\documentclass[aps,prl,twocolumn,groupedaddress]{revtex4-1}
\usepackage{graphicx}

\usepackage{amsmath,scalerel}
\usepackage{xcolor}
\usepackage{subcaption}

\DeclareMathOperator*{\Bigcdot}{\scalerel*{\cdot}{\bigodot}}

\begin{document}

%\preprint{version 0.0}


\title{\textbf{Theoretical study of kinetics of proton coupled electron transfer in photocatalysis}}


\author{Yvelin Giret}
\email[]{yvelin.giret@xmu.edu.cn}
%\thanks{}
%\altaffiliation{1}
\affiliation{State Key Laboratory of Physical Chemistry of Solid Surfaces, iChEM, College of Chemistry and Chemical Engineering,
Xiamen University, Xiamen 361005, China}

\author{Pu Guo}
%\thanks{}
%\altaffiliation{1}
\affiliation{State Key Laboratory of Physical Chemistry of Solid Surfaces, iChEM, College of Chemistry and Chemical Engineering,
Xiamen University, Xiamen 361005, China}

\author{Li-Feng Wang}
%\thanks{}
%\altaffiliation{1}
\affiliation{State Key Laboratory of Physical Chemistry of Solid Surfaces, iChEM, College of Chemistry and Chemical Engineering,
Xiamen University, Xiamen 361005, China}

\author{Jun Cheng}
\email[]{chengjun@xmu.edu.cn}
%\thanks{}
%\altaffiliation{1}
\affiliation{State Key Laboratory of Physical Chemistry of Solid Surfaces, iChEM, College of Chemistry and Chemical Engineering,
Xiamen University, Xiamen 361005, China}


\date{\today}

\begin{abstract}
Photocatalysis induced by sunlight is one of the most promising approach to environmental protection, solar energy conversion and sustainable production of fuels. The computational modeling of photocatalysis is a rapidly expending field which requires to adapt and further develop the available theoretical tools. The coupled transfer of proton and electron is an important reaction during photocatalysis. In this work, we present the first step of our methodology development in which we apply existing kinetic theory of such coupled transfer to a model system, namely, methanol photo-dissociation on rutile TiO$_2$(110) surface, with the help of high-level first-principles calculations. Moreover, we adapt the Stuchebrukhov-Hammes-Schiffer kinetic theory, where we use the Georgievskii-Stuchebrukhova vibronic coupling, to calculate the rate constant of the proton coupled electron transfer reaction for a particular pathway. In particular, we propose a modified expression to calculate the rate constant which enforces the near-resonance condition for the vibrational wavefunction during proton tunneling.
\end{abstract}

%We find that the reaction is vibronically nonadiabatic and in an intermediate regime between electronically adiabatic and electronically nonadiabatic. 

%\keywords{}

\maketitle

\section{I. Introduction}

%\subsection{}
%\subsubsection{}

As more and more societies try to move away from an economy based on fossil fuels and its environmental consequences, clean hydrogen production by solar-driven water splitting appears as a promising direction to follow. Unlocking molecular dihydrogen fuel from water requires a huge amount of energy, and photocatalysis is considered the most reliable process to achieve this goal \cite{van2017perspectives,wang2019recent,hisatomi2019reaction}. Since the pioneering study of Fujishima and Honda in 1972~\cite{fujishima1972electrochemical}, photocatalytic reactions on TiO$_2$ surface has become a widely studied subject \cite{thompson2006surface,FUJISHIMA2008515,henderson2011surface,guo2016elementary}. Shortly after, it has been shown that TiO$_2$ alone was not very active for water splitting, whereas adding sacrificial agents such as methanol could dramatically enhance the efficiency of the reaction \cite{kawai1980photocatalytic}. Despite decades of experimental and theoretical studies of water and methanol photocatalysis on TiO$_2$ surface, the underlying mechanisms are still under debate. A better fundamental understanding of these particular reactions is then of primary importance to help the design of new photocatalysts.

Different scenarios have been proposed in literature to describe the photocatalytic methanol (CH$_3$OH) dissociation on rutile TiO$_2$(110) surface to formaldehyde (CH$_2$O) product, such as direct dehydrogenation \cite{wei2015direct}, implying simultaneous breaking of OH and CH bonds, or a stepwise reaction \cite{feng2016temperature,guo2012stepwise}, in which CH bond cleavage follows OH bond breaking. In all cases, protons are transferred to a bridge oxygen site (O$_{\text{br}}$). It has been argued that chemisorbed methoxy (CH$_3$O) species is first formed by the thermal dissociation of methanol \cite{shen2011identification,zhang2017identifying,chu2016ultrafast}, showing that methoxy, and not molecular methanol, would be the effective hole scavenger. On the other hand, it has been suggested that the first OH breaking step follows an interfacial excitonic proton coupled electron transfer (PCET) mechanism during which the hole is transferred to the adsorbed methoxy species, and that only one photogenerated hole could induce both OH and subsequent CH bond breaking \cite{migani2016excitonic}.

Recently, PCET reactions has been shown to be of general importance in heterogeneous photocatalysis \cite{zhu2017interfacial,chen2015essential,schrauben2012titanium,hoffmann2017proton,chen2013chemical}. Methanol is seen as a hole scavenger during photocatalytic reaction on TiO$_2$ surface which reduces the otherwise high recombination rate of the photo-generated charge carriers, however, the highest occupied molecular orbital (HOMO) of adsorbed species lies below the valence band maximum (VBM) of TiO$_2$, hindering the hole transfer. It has been suggested that the proton transfer to O$_{\text{br}}$ and the hole transfer to methoxy are coupled, the chemical energy required to raise the HOMO being provided by the transferring proton \cite{migani2013level,migani2016excitonic,hao2018photoelectron,migani2014quasiparticle}. From a general point of view, the photoexcited hole can be delocalized (``free'') or localized (``trapped'') at a specific site \cite{di2013hole,ji2014comparative,di2016mechanism}. In particular, it has been shown on the basis of hybrid density functional theory (DFT) calculations on organic adsorbates on anatase TiO$_2$ surface that the molecular adsorption creates a surface dipole which reduces the cost to form a hole at a surface oxygen \cite{di2013hole}. For the particular system under study in this work, it has been shown that HSE06 functional gives correct band alignment, hole localization and PCET thermochemistry~\cite{cheng2012alignment,cheng2014identifying,cheng2015reductive,cheng2014aligning}

In this work, we used density functional theory (DFT) calculations at the hybrid level together with the extension of the Marcus theory to nonadiabatic PCET reactions developed by Hammes-Schiffer, Soudackov, and co-workers \cite{hammes2010theory,ghosh2017theoretical} to study the first step of methanol photo-dissociation on rutile TiO$_2$(110) surface. The manuscript is organized as follows: In section II we briefly present the theoretical approach used and the details of the first-principles calculations we performed. In section III we present our results and discuss them, and we conclude in section IV. Finally, we present in the appendices some details of the calculations realized to obtain the different parameters.

\section{II. Methodology}

\subsection{1. Vibronic Coupling and Rate Constant}

The form of the PCET rate constant will principally depend on the value of the vibronic couplings $V_{\mu\nu}$ between the reactant ($\mu$) and product ($\nu$) electron-proton vibronic states, defined as the Hamiltonian matrix element between the reactant and product electron-proton vibronic wave functions. In the vibronically adiabatic limit, where $V_{\mu\nu} \gg k_{_{\text{B}}}T$, the rate constant can be estimated by the transition state theory (TST) which will be essentially independent of the vibronic couplings:
\begin{equation}\label{k_tst}
k_{_{\text{TST}}} \propto \left( \frac{k_{_{\text{B}}} T}{2 \pi \hbar} \right) \exp \left[ - \frac{\Delta G^{\ddagger}}{ k_{_{\text{B}}} T} \right].
\end{equation}
where $\hbar$ is the reduced Planck's constant, $k_{_{\text{B}}}$ the Boltzmann's constant, $T$ the absolute temperature, and $\Delta G^{\ddagger}$ the free energy of activation. In the vibronically nonadiabatic limit, where $V_{\mu\nu} \ll k_{_{\text{B}}}T$, the PCET rate constant can be calculated by the Stuchebrukhov-Hammes-Schiffer (SHS) theory \cite{hammes2008proton,hammes2010theory}, where the rate constant for coupled electron and proton transferring together is written as an extension of the Marcus theory for electron transfer:
\begin{eqnarray}\label{k_shs}
k_{_{\text{SHS}}} &=&  \sum_{\mu} P_{\mu} \sum_{\nu} \frac{\left|V_{\mu\nu}\right|^2}{\hbar} \sqrt{\frac{\pi}{\lambda_{\mu\nu} k_{_{\text{B}}}T}} \nonumber \\
&\times& \exp \left[ - \frac{(\Delta G_{\mu\nu} + \lambda_{\mu\nu})^2}{4 \lambda_{\mu\nu} k_{_{\text{B}}} T}  \right],
\end{eqnarray}
where the double summation is over all pairs of reactant ($\mu$) and product ($\nu$) electron-proton vibronic states, $P_{\mu}$ is the Boltzmann population of the reactant state $\mu$, $\Delta G_{\mu\nu}$ is the reaction free energy, and $\lambda_{\mu\nu}$ is the reorganization energy.

%%%%%%%%%%%%%%%%%%%%%%%%%%%%%%%%%%%%%%%%%%%%%%%%%%%%%%%%
\begin{figure}[ht!]
\vskip0.5truecm
\begin{center}
\includegraphics[angle=0,width=0.5\textwidth]
{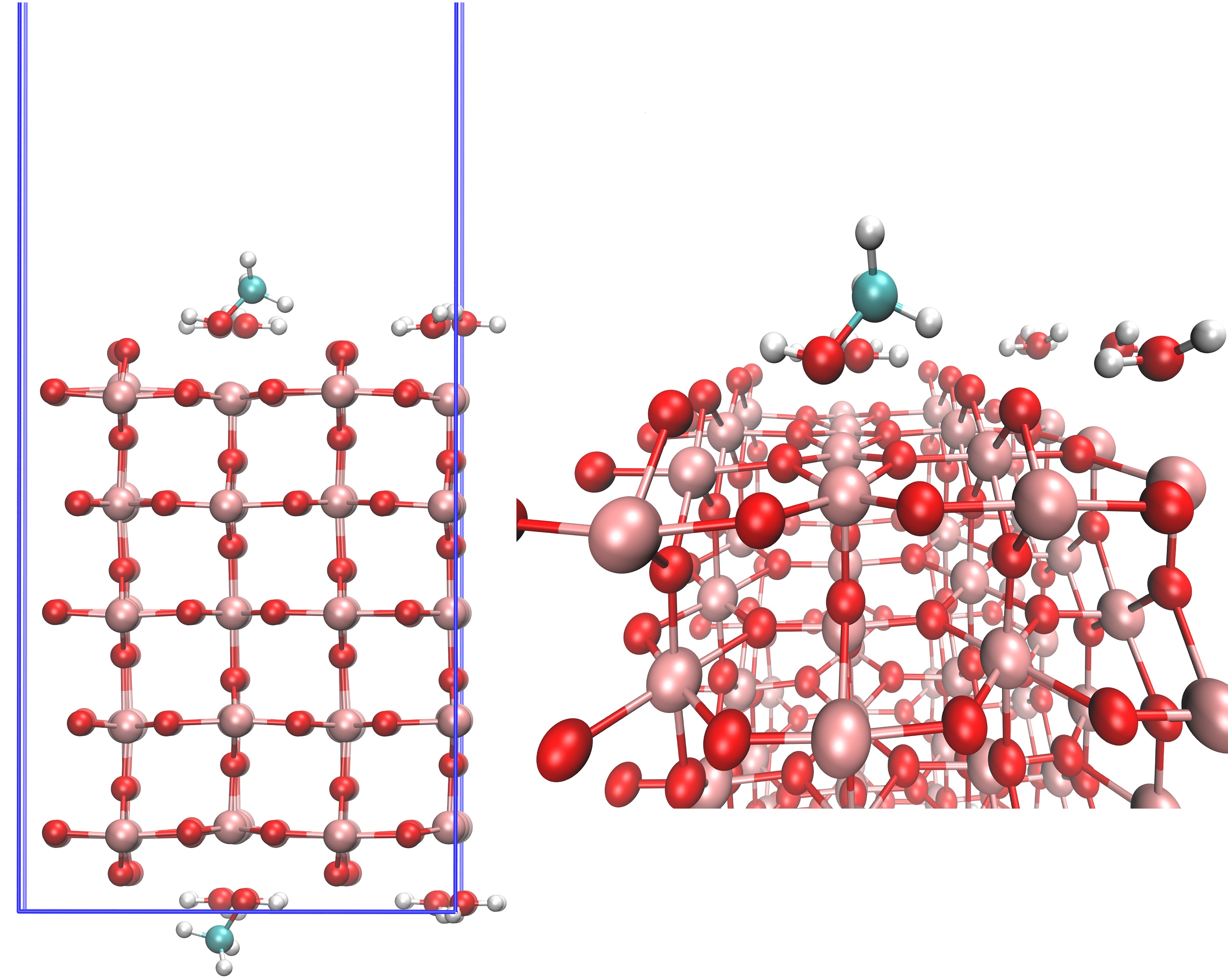}
\end{center}
\caption{
\label{struc-init}
Initial structure with one monolayer on each side of the slab, each one containing 7 water molecules and 1 methanol molecule (see text). Left: view from the side with the supercell edges (orthographic). Right: zoom on the top part (perspective). The methanol molecule adsorbs directly on the TiO$_2$(110) surface. Pink: Ti, Red: O, Green: C, white: H.}
\end{figure}
%%%%%%%%%%%%%%%%%%%%%%%%%%%%%%%%%%%%%%%%%%%%%%%%%%%%%%%%%

%%%%%%%%%%%%%%%%%%%%%%%%%%%%%%%%%%%%%%%%%%%%%%%%%%%%%%%%%%%%%%%%%%%%%%%%%%%%%%%%%%%%%%%%%
\begin{figure}[h]
\centering
 \begin{tabular}{c c}
\includegraphics[width=0.45\linewidth]{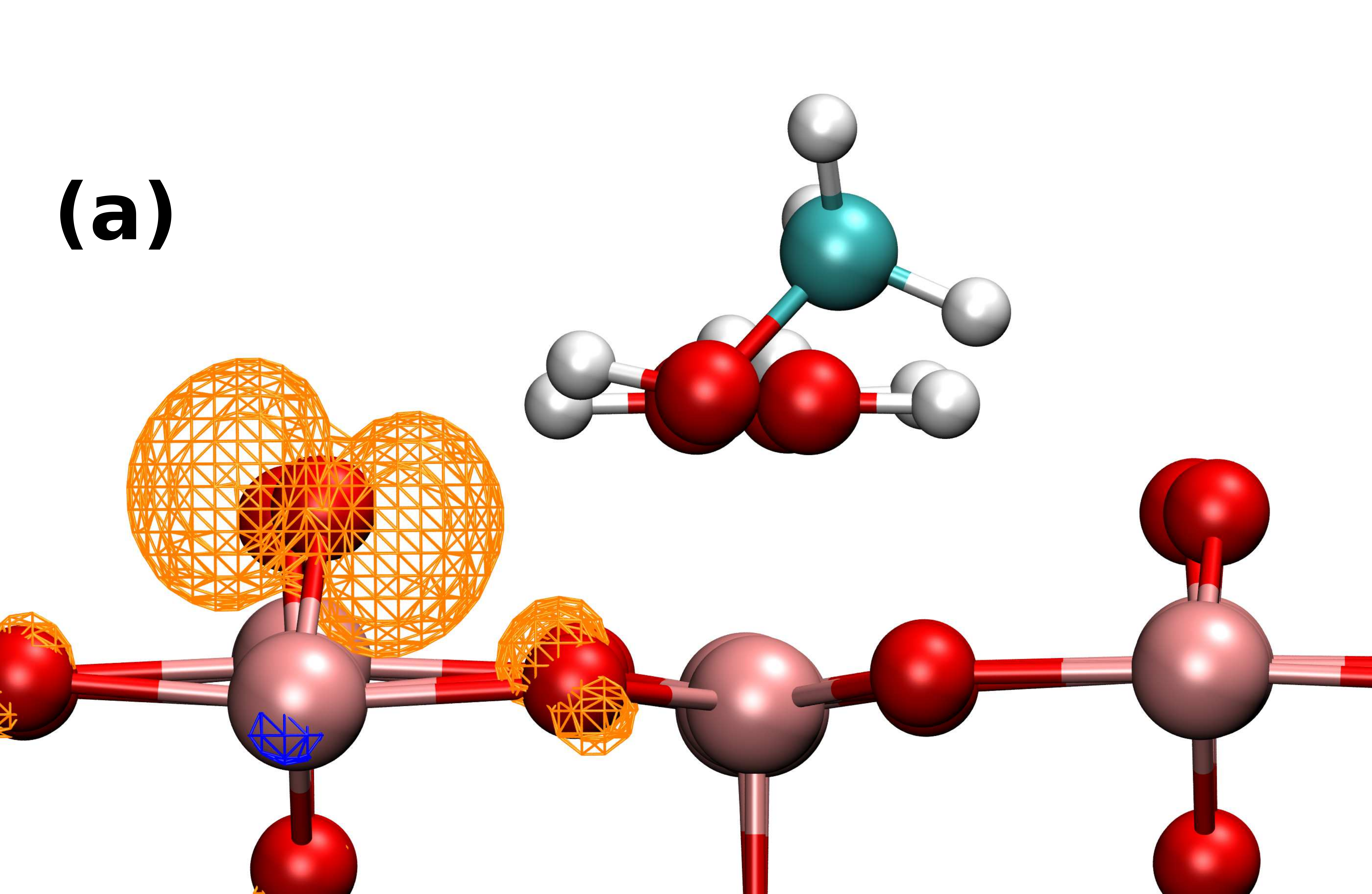} & \includegraphics[width=0.45\linewidth]{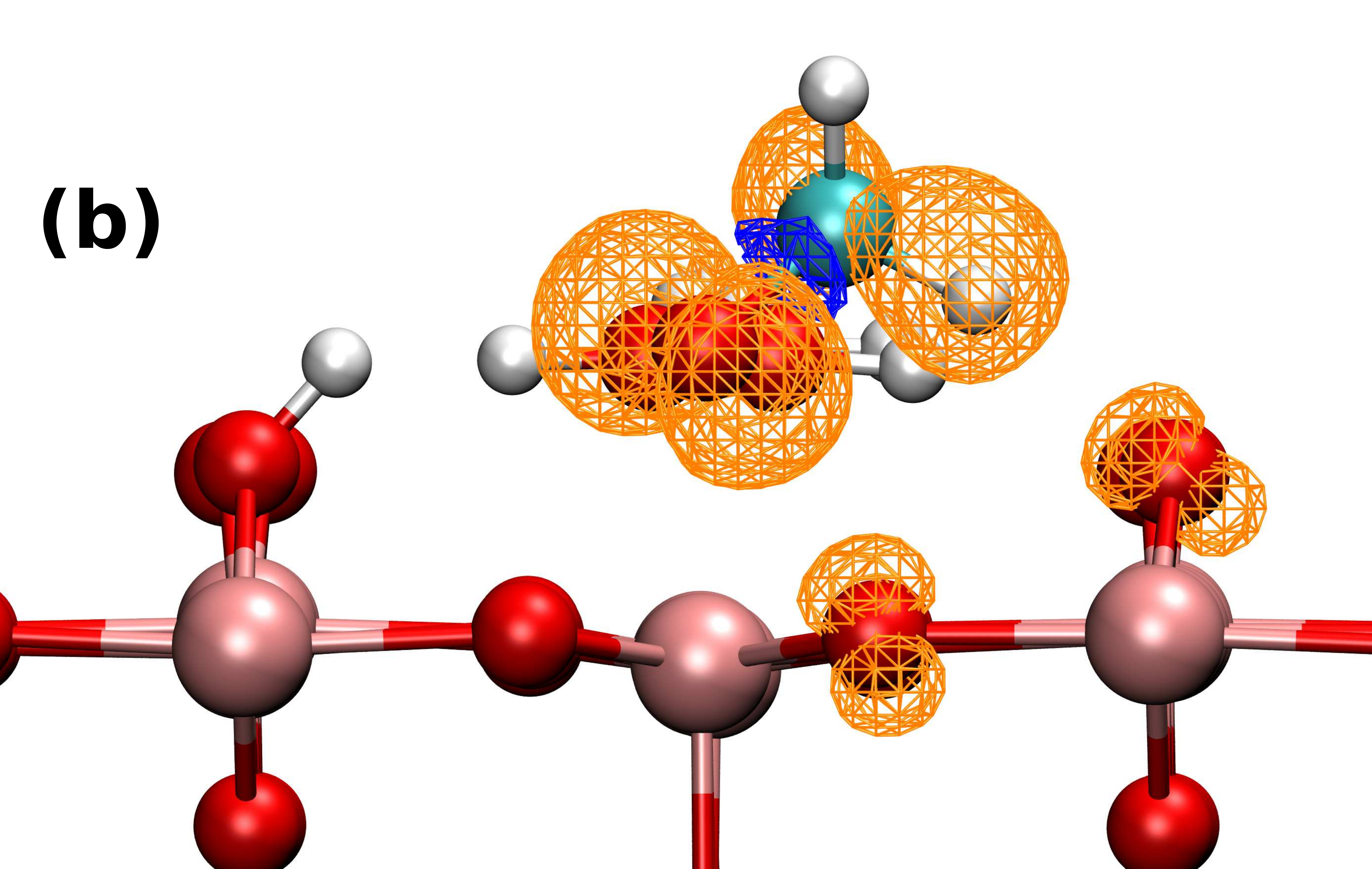} \\
%\vspace{-0.8cm}
\includegraphics[width=0.45\linewidth]{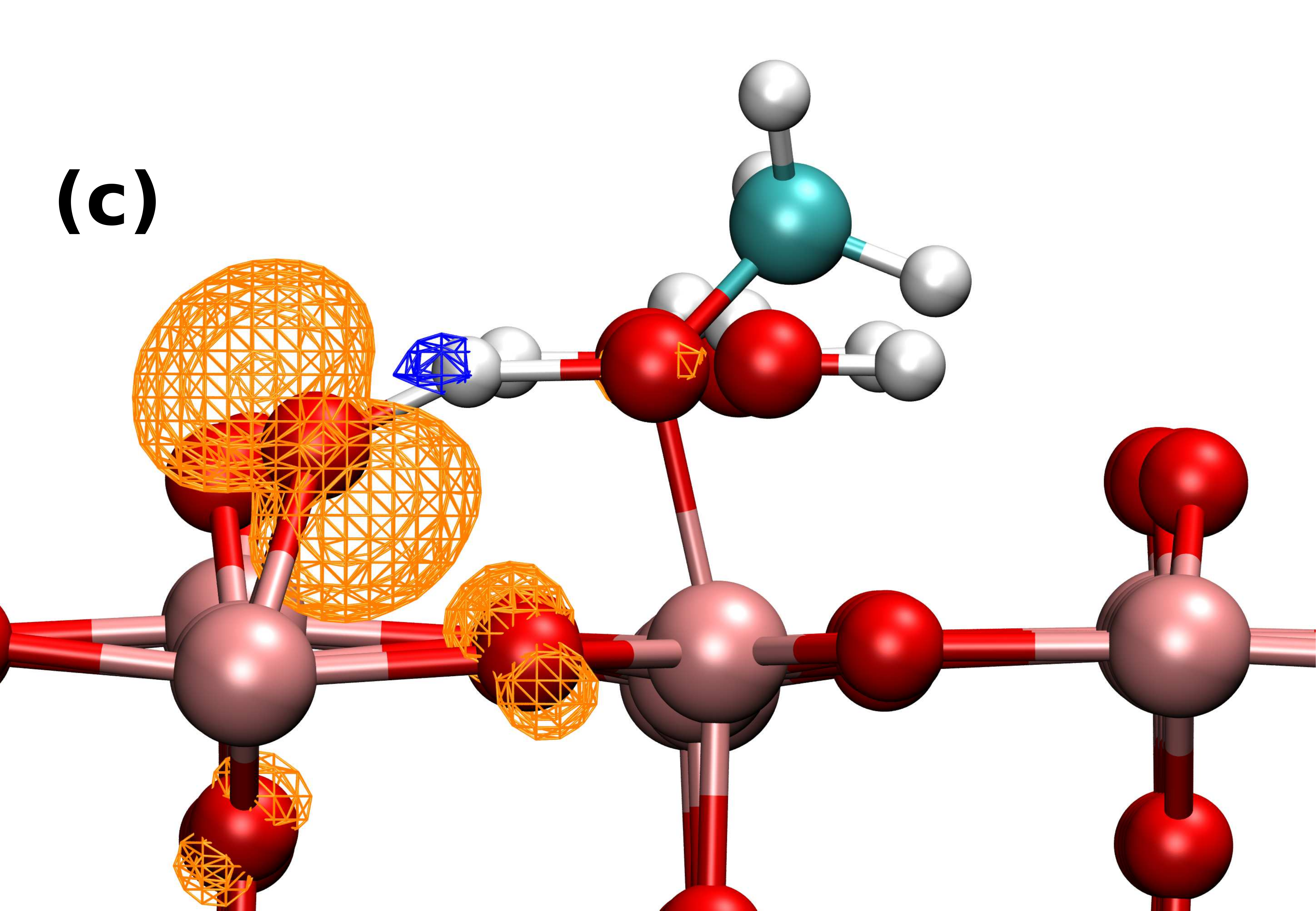} & \includegraphics[width=0.45\linewidth]{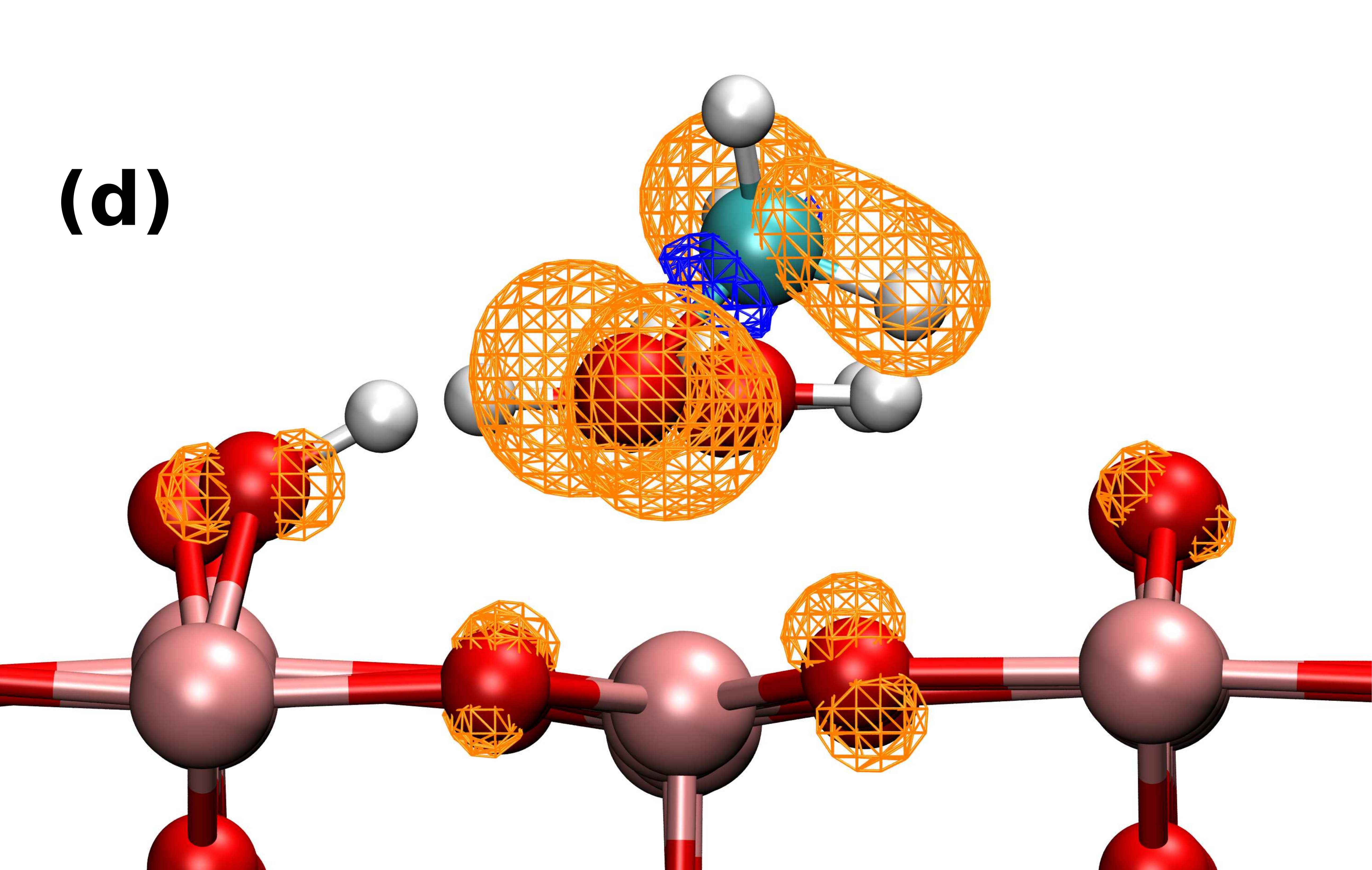} \\
\includegraphics[width=0.45\linewidth]{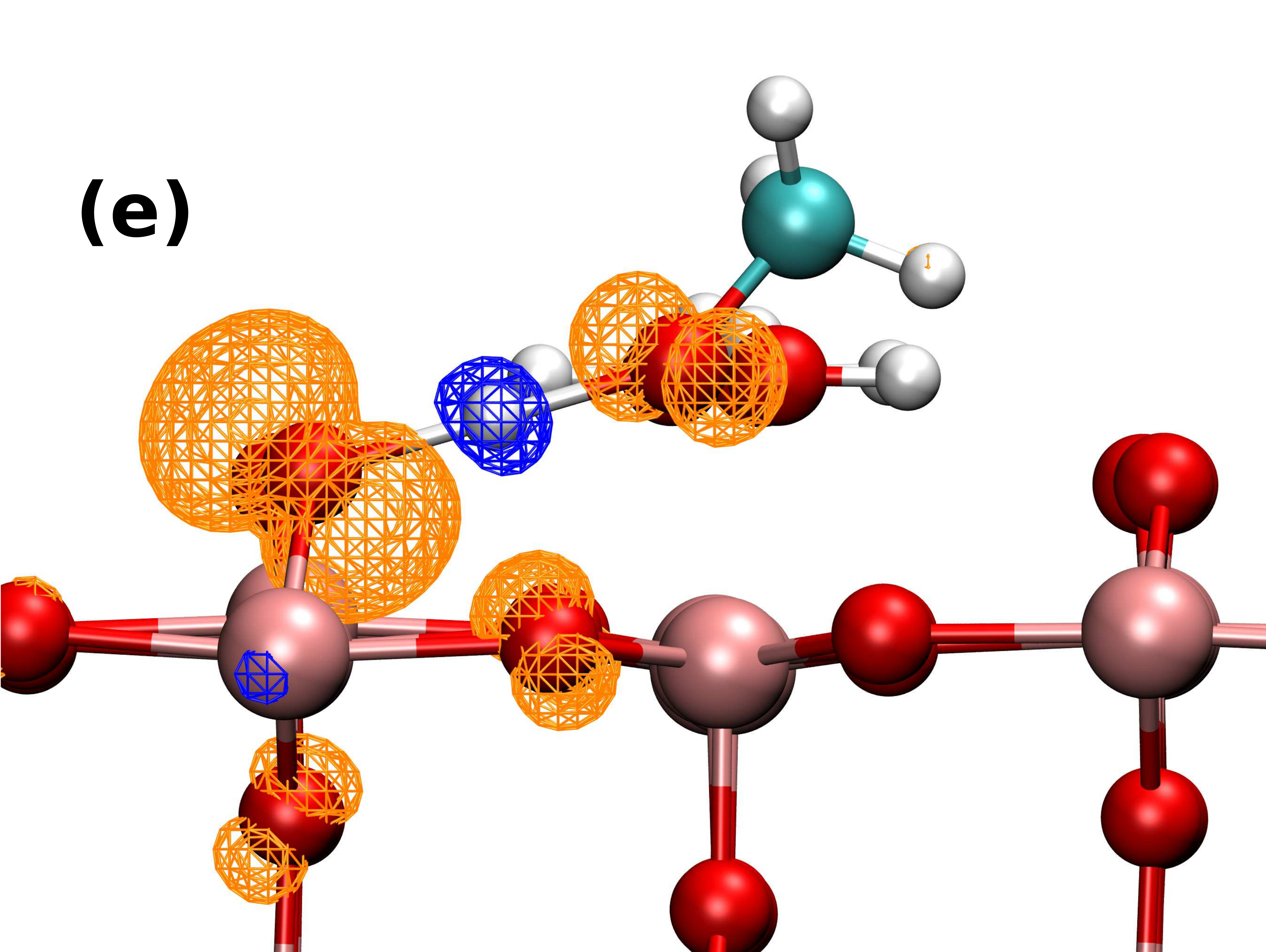} & \includegraphics[width=0.45\linewidth]{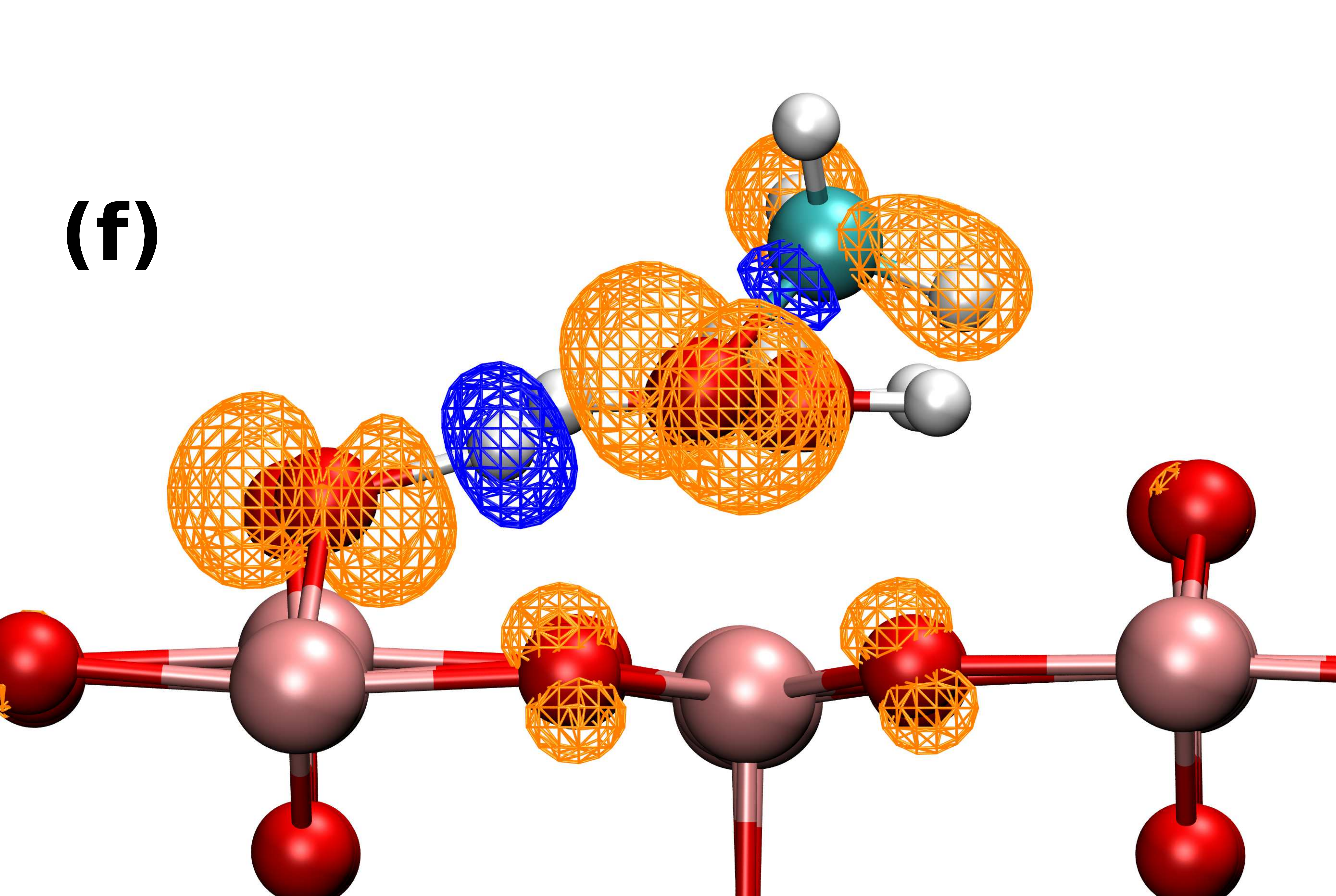} \\
\end{tabular}
\caption{Spin densities calculated with HSE06 functional including a hole. (a): reactant state, (b): product state, (c): transition state, (d): NEB state after the transition state, (e) and (f) are two states along the proton potential calculated in the reactant state geometry. (Orange: positive. Blue: Negative.)}
\label{spin-densities}
\end{figure}
%%%%%%%%%%%%%%%%%%%%%%%%%%%%%%%%%%%%%%%%%%%%%%%%%%%%%%%%%%%%%%%%%%%%%%%%%%%%%%%%%%%%%%%%%

The vibronic coupling can be estimated by the semiclassical (sc) approach given by Georgievskii and Stuchebrukhova \cite{georgievskii2000concerted} which spans the two limiting cases of small and large electronic coupling. For small electronic coupling, the reaction will be electronically nonadiabatic and the electron will not respond instantaneously to the proton motion, whereas for large electronic coupling the reaction will be electronically adiabatic. The general form of the vibronic coupling is given by \cite{georgievskii2000concerted}:
\begin{equation}\label{vib_sc}
V^{(\text{sc})}_{\mu\nu} = \kappa V^{(\text{ad})}_{\mu\nu},
\end{equation}
where $V^{(\text{ad})}$ is the vibronic coupling in the electronically adiabatic limit, and $\kappa$ is the factor assessing the electronic nonadiabaticity of the reaction:  
\begin{equation}\label{kappa}
\kappa = \sqrt{2\pi p} \frac{e^{p\ln p-p}}{\Gamma(p+1)},
\end{equation}
where $\Gamma(x)$ is the gamma function, and $p$, called adiabatic parameter, is given by:
\begin{equation}\label{p}
p = \frac{| H_{_{\text{DA}}} |^2}{ \hbar |\Delta F|v_t} = \frac{\tau_p}{\tau_e},
\end{equation}
where $H_{_{\text{DA}}}$ is the electronic coupling between the initial (donor) and final (acceptor) states, as appearing in the Marcus theory, $\Delta F$ is the difference between the slopes of the proton potential energy curves at the crossing point, $v_t = (2(V_c-E)/m)^{1/2}$ is the ``tunneling velocity'' of the proton at the crossing point, with $m$ the mass of the proton, $V_c$ the potential energy at the crossing point, and $E$ the tunneling energy. $\tau_p = | H_{_{\text{DA}}} |/|\Delta F|v_t$ and $\tau_e = \hbar / H_{_{\text{DA}}}$ are respectively the proton ``tunneling time'' and the time required to change the electronic state~\cite{georgievskii2000concerted}.

For $p \ll 1$ ($\kappa = \sqrt{2\pi p}$), the reaction will be electronically nonadiabatic (na), and the vibronic coupling will be reduced to:
\begin{equation}\label{vib_na}
V^{(\text{na})}_{\mu\nu} = | H_{_{\text{DA}}} | \langle \varphi_{\mu} | \varphi_{\nu} \rangle,
\end{equation}
where $\varphi_{\mu,\nu}$ are the proton vibrational wave functions. For $p \gg 1$ ($\kappa = 1$), the reaction will be electronically adiabatic and the vibronic coupling will be given by \cite{georgievskii2000concerted}:
\begin{equation}\label{vib_ad}
V^{(\text{ad})}_{\mu\nu} = \hbar c_{\mu} c_{\nu} \sqrt{\omega_{_{\text{D}}}\omega_{_{\text{A}}}} \exp \left[ - \int_{x_{\mu}}^{x_{\nu}} k(x')\text{d}x'\right],
\end{equation}
where $x_{\mu}$ and $x_{\nu}$ delimit the tunneling region, $\omega_{_{\text{D}}}$ and $\omega_{_{\text{A}}}$ are the vibration frequencies in the donor and the acceptor wells. $c_{\mu}$ and $c_{\nu}$ are numerical coefficients corresponding to the $\mu^{\text{th}}$ and $\nu^{\text{th}}$ excited states of the proton respectively in the donor and acceptor wells:
\begin{equation}
c_n = \pi^{-1/4} \sqrt{\frac{2^n}{n!}} \exp \left[  \frac{\left(2n+1\right)\left(\ln(2n+1)-2\ln 2 -1\right)}{4} \right].
\end{equation}
The function $k(x)$ appearing in Eq. (\ref{vib_ad}) is the wave vector of the wave function written in the Wentzel-Kramers-Brillouin (WKB) approximation \cite{nitzan2006chemical}:
\begin{equation}\label{wkb}
k(x) = \frac{\sqrt{2m(V(x)-E)}}{\hbar},
\end{equation}
where $V(x)$ is the proton potential (see below).

To enforce the tunneling near-resonance condition between the proton vibrational states, we modified Eq. (\ref{k_shs}), and the rate constant is calculated from: 
\begin{eqnarray}\label{k_me}
k_{_{\text{PCET}}} &=&  \sum_{\mu} P_{\mu} \sum_{\nu} \frac{\left|V_{\mu\nu}\right|^2}{\hbar} \sqrt{\frac{\pi}{\lambda k_{_{\text{B}}}T}} \exp \left[ - \frac{\left| E_{\nu} - E_{\mu} \right|}{k_{_{\text{B}}} T}  \right] \nonumber \\
&\times& \exp \left[ - \frac{(\Delta G^0 + \lambda)^2}{4 \lambda k_{_{\text{B}}} T}  \right], 
\end{eqnarray}
where $\Delta G^0$ is the free energy of reaction as appearing in the Marcus theory, $E_{\mu/\nu}$ are the proton vibrational energy levels, and where the reorganization energy $\lambda$ is calculated once for the whole reaction. Details on the calculation of the different quantities appearing in equation (\ref{k_me}) are given in the appendices. Briefly, the electronic coupling $H_{_{\text{DA}}}$ is calculated by the projection-operator diabatization (POD) method \cite{kondov2007quantum,futera2017electronic} (see appendix B). The electron-proton vibronic states $(\mu/\nu)$ are calculated by fitting the \textit{ab initio} potential energy curves by Morse potentials (see appendix A). And the reorganization energy is calculated by a variant of the four-point model \cite{ghosh2017theoretical,fernandez2012theoretical} (see appendix C).

\subsection{2. Computational Details}

%%%%%%%%%%%%%%%%%%%%%%%%%%%%%%%%%%%%%%%%%%%%%%%%%%%%%%%%%%%%%%%%%%%%%%%%%%%%%%%%%%%%%%%%%
\begin{figure*}[ht!]
\centering
 \begin{tabular}{c c c }
\includegraphics[width=0.33\linewidth]{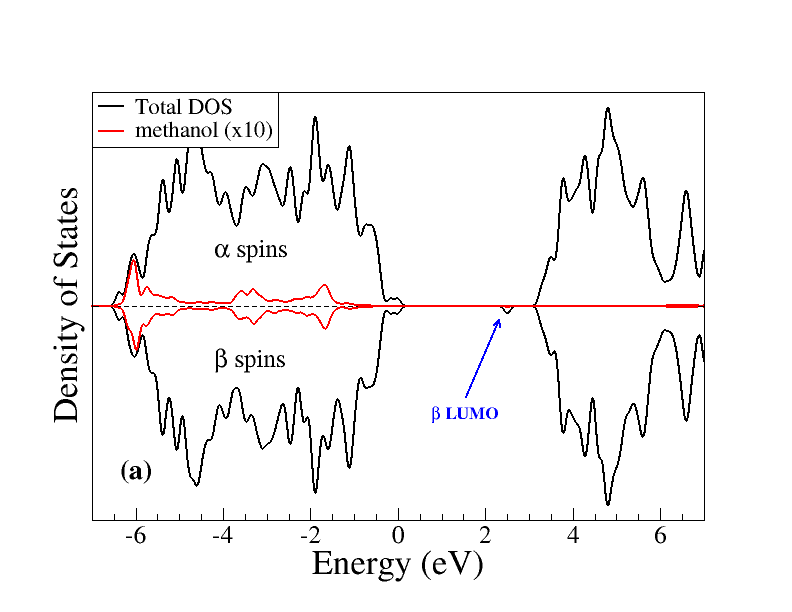} & \includegraphics[width=0.33\linewidth]{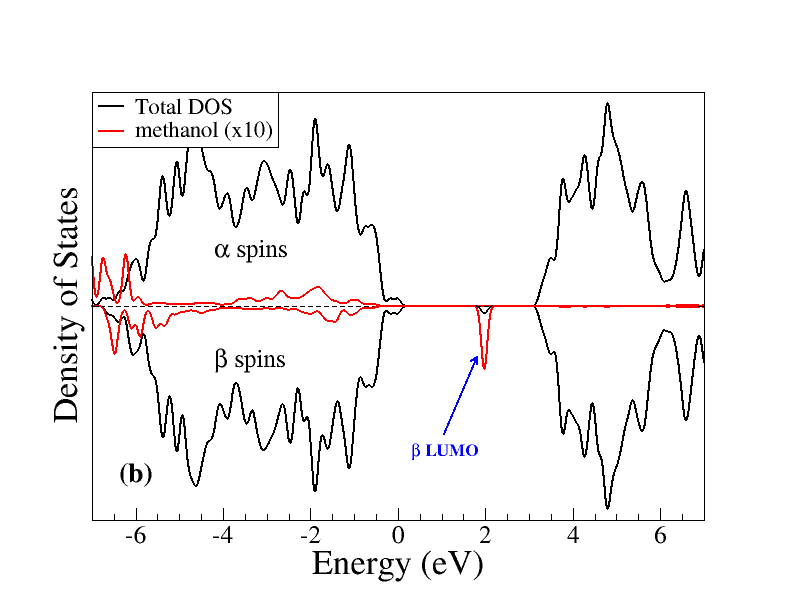} & \includegraphics[width=0.33\linewidth]{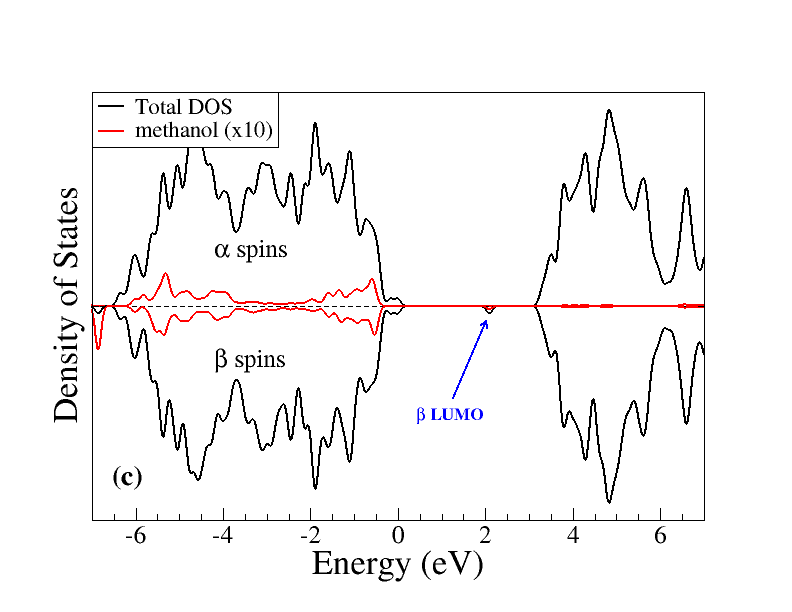} \\
%\vspace{-0.8cm}
\includegraphics[width=0.33\linewidth]{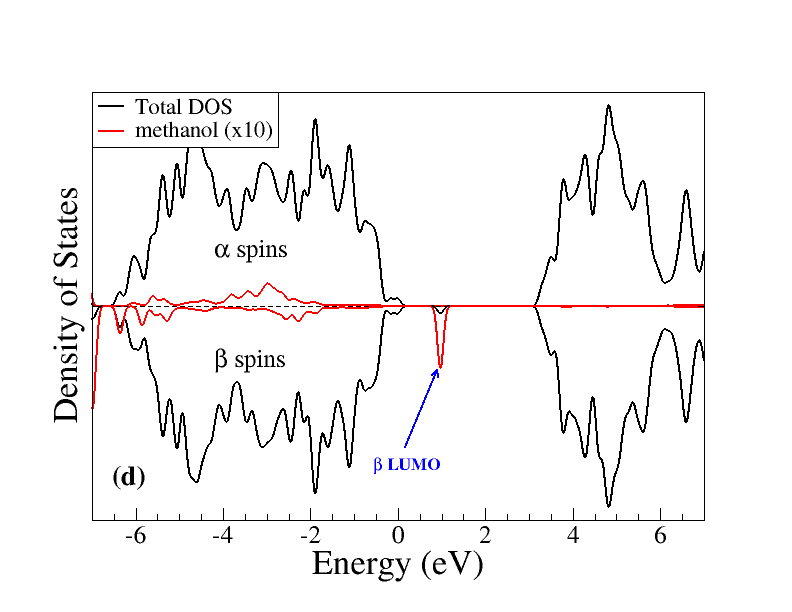} & \includegraphics[width=0.33\linewidth]{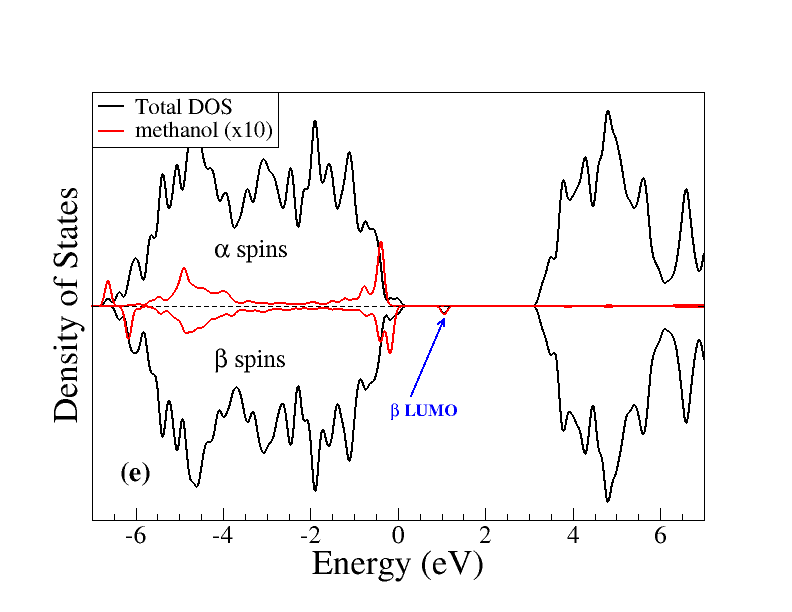} & \includegraphics[width=0.33\linewidth]{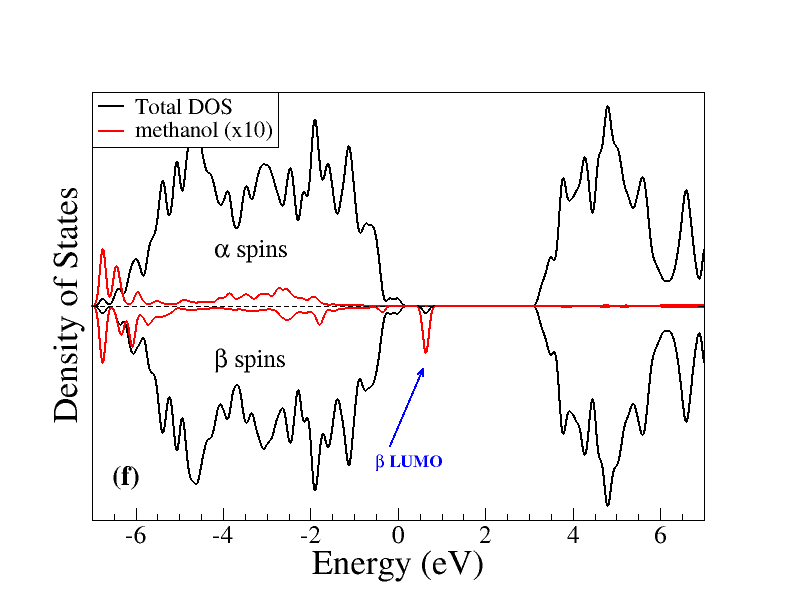} \\
\end{tabular}
\caption{Densities of states (black: total DOS, red: projected DOS) for the structures shown in Fig. \ref{spin-densities}. (a): reactant state, (b): product state, (c): transition state, (d): NEB state after the transition state, (e) and (f) are two states along the proton potential calculated in the reactant state geometry.}
\label{dos}
\end{figure*}
%%%%%%%%%%%%%%%%%%%%%%%%%%%%%%%%%%%%%%%%%%%%%%%%%%%%%%%%%%%%%%%%%%%%%%%%%%%%%%%%%%%%%%%%%

The rutile TiO$_2$(110) surface was modeled by a five O-Ti-O trilayers slab with the $xy$ dimensions of 4$\times$2 supercell, as shown in Fig.~\ref{struc-init}, giving a TiO$_2$ slab with 240 atoms and a supercell containing 294 atoms in total. We used experimental lattice parameters at 295~K: $a=4.593$~\AA~and $c=2.959$~\AA~\cite{burdett1987structural}. 3D periodic boundary conditions are used with a $\sim$20~\AA~vacuum gap between periodic image in the $z$ direction (corresponding to the (110) direction), giving a supercell of dimensions 11.836$\times$12.991$\times$35.0 \AA$^3$. We placed one monolayer (ML) of water molecules on the surface, consisting of 8 water molecules on unsaturated Ti sites, for which the orientation have been chosen such as to maximize Hydrogen bonding and symmetries. Moreover, to avoid the building of a net dipole moment, we placed a symmetric layer at the bottom of the slab. The thickness of the slab was investigated to ensure convergence on band positions and water adsorption energies. Similar setup has been used for extensive tests in previous studies published by our group \cite{cheng2015reductive,cheng2010acidity,cheng2010aligning,cheng2012hole}. Finally, one water molecule is replaced by one methanol molecule on each side of the slab, which correspond to a coverage of~1/8~ML.

All \textit{ab initio} calculations were performed using the freely available CP2K/Quickstep software package \cite{vandevondele2005quickstep,hutter2014cp2k}. The core electrons were represented by analytical Goedecker-Teter-Hutter (GTH) pseudopotentials \cite{hartwigsen1998relativistic}. For Ti atoms, the explicit valence electrons are 3$s^2$3$p^6$4$s^2$3$d^2$, for O atoms 2$s^2$2$p^4$, for C atoms 2$s^2$2$p^2$, and for H atoms 1$s^1$. The basis sets were short-ranged (less diffuse) double-$\zeta$ basis functions with one set of polarization functions (DZVP) \cite{vandevondele2007gaussian}. We also tested the adsorbed molecules in gas phase using larger basis sets (TZV2P) and found a negligible difference of 0.01 eV in the adsorption energies. After convergence study, the plane wave cutoff for the electron density was fixed to 500 Ry. Given the size of the supercell, all calculations were done at the $\Gamma$ point only. Geometry optimization was done with the help of the Broyden-Fletcher-Goldfarb-Shanno (BFGS) algorithm allowing all the atoms to relax until the forces were down to 4.5$\times$10$^{-4}$~Ha/bohr.

Standard LDA or GGA functionals have the tendency to over-delocalize electronic states in transition metal oxides such as TiO$_2$ \cite{cheng2015reductive}. Two approaches are possible to limit this problem, namely, the use of hybrid functionals (which include partial exact Hartree-Fock exchange) or the use of DFT+$U$ calculations (which include on-site Coulombic repulsion) \cite{kulik2015perspective,wang2015identifying}. Here, we used a Heyd-Scuseria-Ernzerhof (HSE) hybrid functional in the HSE06 version \cite{heyd2003hybrid,krukau2006influence}, for which the calculations are carried out applying an auxiliary density matrix method (ADMM) recently developed and implemented in CP2K \cite{guidon2010auxiliary,guidon2009robust,guidon2008ab}. In this method, the density matrix is re-expanded in a small auxiliary basis set leading to massive speed-up of the calculation of Hartree Fock exchange. We included van der Waals interaction through the use of a Grimme's dispersion correction (DFT-D3) \cite{grimme2010consistent}. All calculations were spin-polarized.

\subsection{3. Model Validation}

We first validate our implementation by reproducing some of the results published recently by Ghosh \textit{et al.} on adsorbed organic radical (TEMPO) on a photoreduced ZnO nanocrystal \cite{ghosh2017theoretical}. We refer the reader to Ref. \cite{ghosh2017theoretical} for the details about this system. From their calculations, the reorganization energy was calculated to be 1.6~eV from the approach given by Eq. (\ref{lambda_in} in the Appendices), with a similar contribution from TEMPO and ZnO components. From their geometries optimized with a PBE functional, we have calculated the energies with the hybrid functional PBE0-TC-LRC \cite{guidon2009robust}, and obtain 0.717~eV for TEMPO and 0.895~eV for the ZnO nanocrystal, giving a total inner-sphere reorganization energies of 1.612~eV, in good agreement with the published result \cite{ghosh2017theoretical}. We also did the same calculation with optimized structure with the PBE0-TC-LRC functional, and obtained 0.683~eV for TEMPO and 0.916~eV for the ZnO nanocrystal, giving a total inner-sphere reorganization energies of 1.599~eV. Finally, we did the same calculation with the HSE06 functional, and obtained 0.689~eV for TEMPO and 0.935~eV for the ZnO nanocrystal, giving a total inner-sphere reorganization energies of 1.624~eV. We also compared the proton wavefunction overlap at a donor-acceptor distance of 2.30~\AA~through the use of Eq. (\ref{proton_WF}), calculated to be $S_{00}^2 = 1.213\times10^{-2}$ and  $S_{01}^2 = 9.597\times10^{-2}$, in excellent agreement with Ref.~\cite{ghosh2017theoretical}.

We have calculated the electronic coupling by implementing the POD method (see appendix B) for a charged (+1) He dimer for a DZVP basis set and a PBE50 functional (50~\% Hartree-Fock exchange) as recently published by Futera and Blumberger \cite{futera2017electronic}. For an interatomic distance of respectively 2.5~\AA, 3.0~\AA, 4.0~\AA, and 5.0~\AA, we calculated (from Ref. \cite{futera2017electronic}) 194.91~meV (196.22~meV), 62.74~meV (64.55~meV), 5.70~meV (6.30~meV), and 0.28~meV (0.64~meV), which show a reasonable agreement. These different results give us confidence in our implementation, which we now use to describe photo-dissociation of methanol adsorbed on rutile TiO$_2$(110) surface.

\section{III. Results and Discussion}

We first optimized the initial structure shown in Fig.~\ref{struc-init} including a hole. To ensure hole localization at the surface, we slightly modified the position of the nearest oxygen atom to the transferring proton before optimization, as the hole localization site will be determined by the starting structure used in geometry optimization procedure~\cite{cheng2015reductive}. The initial spin density will then be localized preferentially on O$_{2p}$ orbitals, as shown in Fig. ~\ref{spin-densities} (a). We then optimized the product structure geometry by moving a proton from adsorbed molecule to the nearest oxygen atom, as shown in Fig.~\ref{spin-densities} (b), where the hole is now localized on the dissociated molecule. The corresponding total and projected density of states are shown in Fig. \ref{dos}, where we can see that the trapped holes in the initial structure have a vertical energy level inside the band gap well above the VBM at 2.50~eV, in agreement with previous calculations at the hybrid level on the rutile TiO$_2$(110) water interface \cite{cheng2014identifying}. In the final structure, the hole is localized on the adsorbed species with a vertical energy level inside the bandgap at 1.98~eV above the VBM. From a general point of view, these results tend to confirm that the transferring proton will raise the HOMO of the adsorbed species and promote the hole transfer.

\begin{figure}[t]
\vskip0.5truecm
\begin{center}
\includegraphics[angle=0,width=0.5\textwidth]
{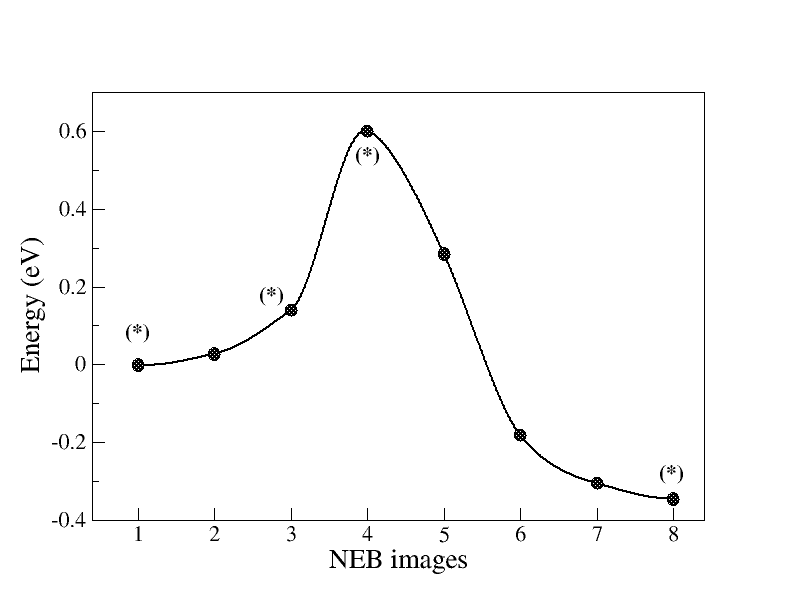}
\end{center}
\caption{
\label{NEB-paths}
Adiabatic energy profiles including a hole obtained from CI-NEB with an HSE06 functional. (*) indicates a geometry at which the proton potential $V(x)$ is calculated.  }
\end{figure}

From these structures we obtain the Gibbs free energy of reaction $\Delta G^0$, which is calculated to be $-0.34$~eV, showing that the reaction is highly exothermic. We then used the climbing image nudged elastic band (CI-NEB) method \cite{henkelman2000climbing} to generate the reaction path for proton transfer from methanol to titania surface and identify the transition state (TS), as shown in Fig. \ref{NEB-paths}.
%%%%%%%%%%%%%%%%%%%%%%%%%%%%%%%%%%%%%%%%%%%%%%%%%%%%%%%%
%%%%%%%%%%%%%%%%%%%%%%%%%%%%%%%%%%%%%%%%%%%%%%%%%%%%%%%%%
The adiabatic energy barrier ($\Delta G^{\ddagger}$) is calculated to be 0.60 eV, in good agreement with calculations recently performed at the DFT$+U$ level by Zhang \textit{et al.} \cite{zhang2017identifying}. No stable intermediates are observed, suggesting that the reaction path involved the coupled transfer of an electron and a proton. We then assume that OH bond breaking is a concerted PCET mechanism and use the methodology presented in section II.1.

The electronic coupling is calculated at the transitions state based on the projection-operator diabatization (POD) method \cite{kondov2007quantum,futera2017electronic} (see appendix B). The system is separated between a hole donor part (D), containing TiO$_2$ atoms, the water molecules and the bottom layer methanol molecule, and a hole acceptor part (A), containing the active methanol molecule. The molecular orbitals corresponding to the spin density at and after the TS are used to calculate the electronic coupling. We choose to keep the transferring proton in the donor part as this partitioning scheme preserves the shape of the orbitals of interest (in particular the hole state), allowing us to identify the initial and the final states of the hole transfer. The electronic coupling is then calculated to be 0.186~eV, suggesting an intermediate regime between electronically adiabatic and nonadiabatic. Using Eq. (\ref{lambda_in}), the reorganization energy is calculated to be 0.091~eV. The value of reorganization energy can be divided in two parts describing the geometrical reorganization of the surface of the semiconductor and in the molecule, which are calculated to be 41 and 50 meV respectively, indicating a similar contribution. Such small reorganization energy is typical of reactions which do not involved an important charge rearrangement.

\begin{figure*}
\centering
 \begin{tabular}{c c}
\includegraphics[width=0.45\linewidth]{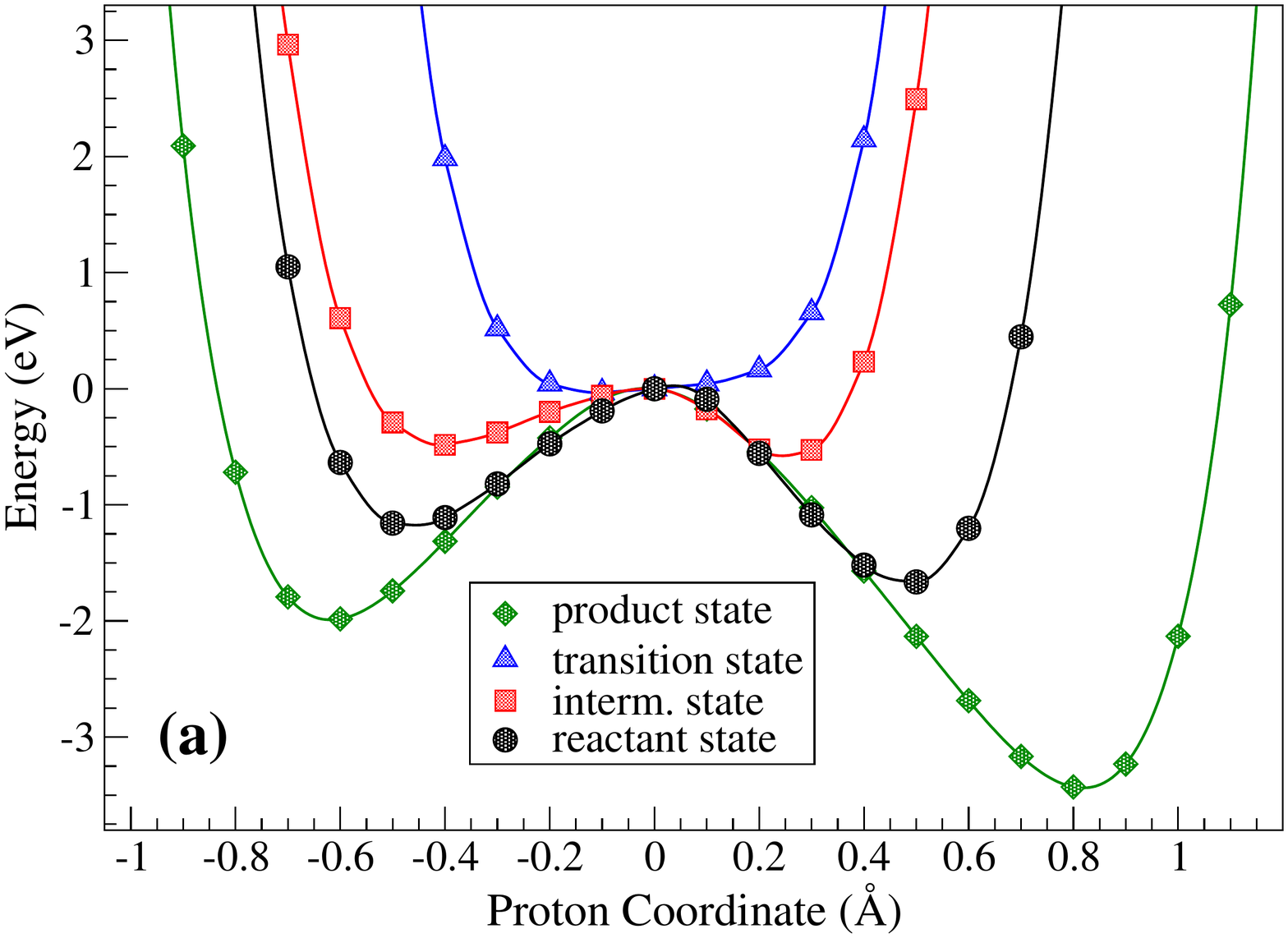} & \includegraphics[width=0.45\linewidth]{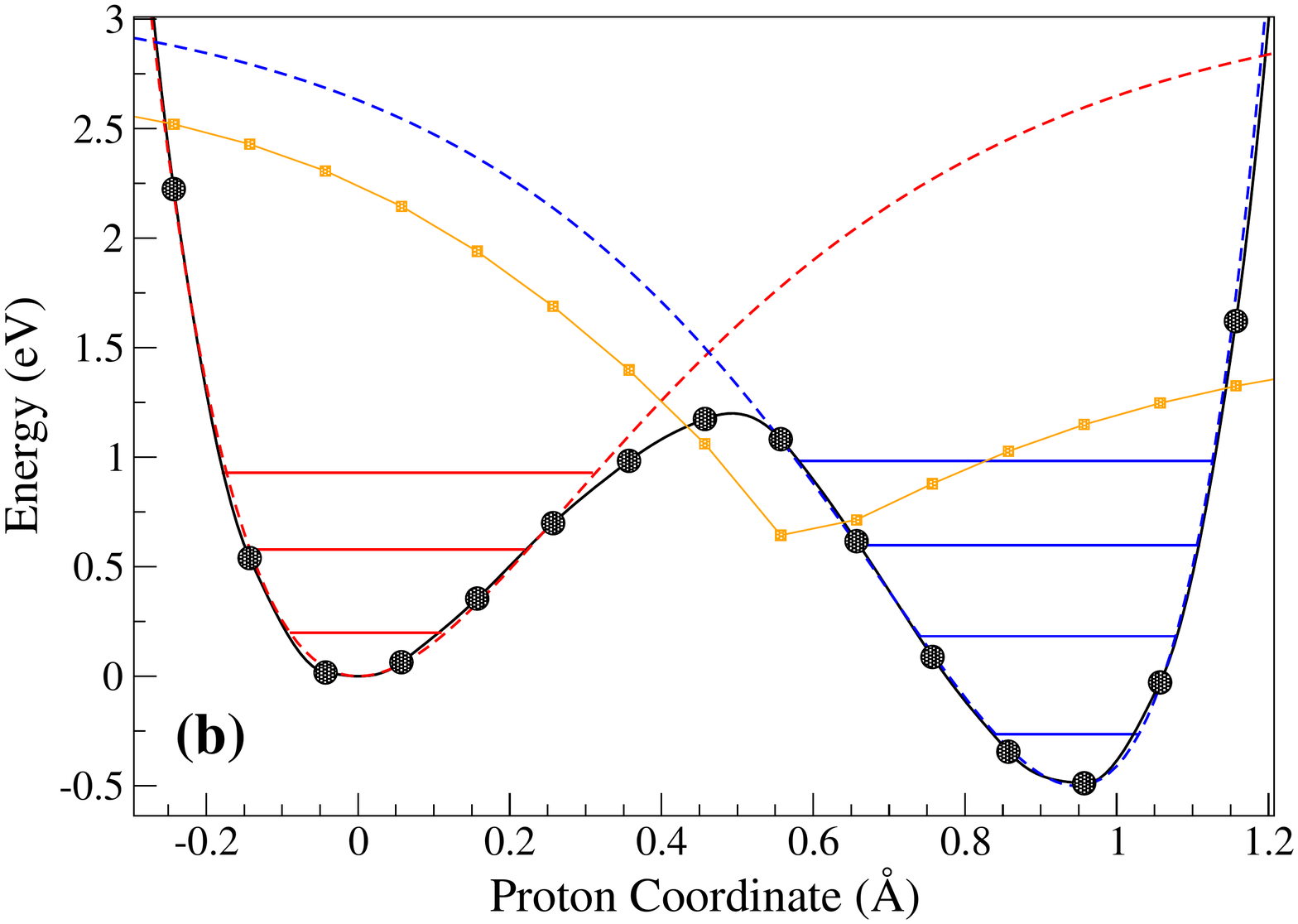} \\
%\vspace{-0.8cm}
\includegraphics[width=0.45\linewidth]{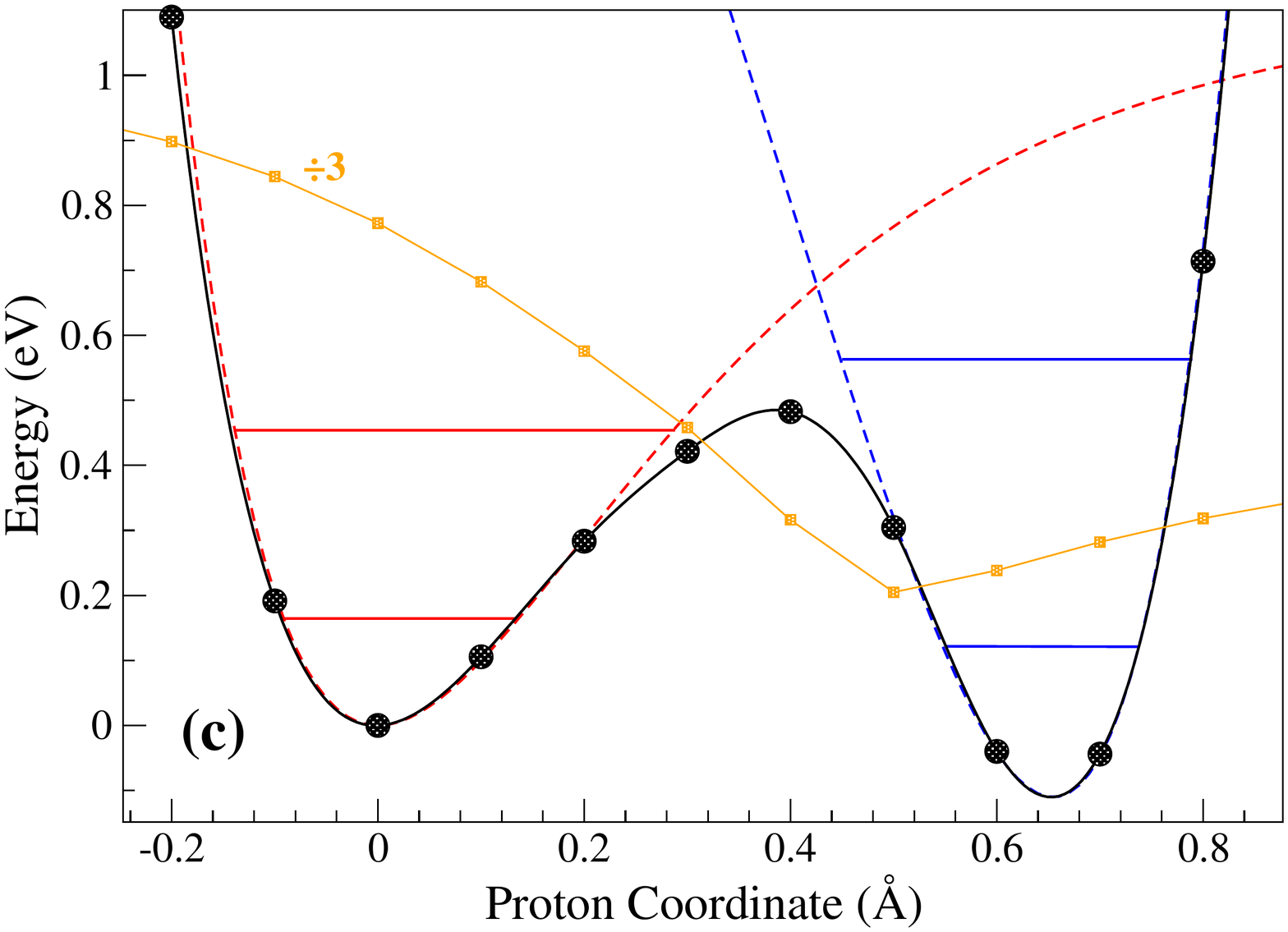} & \includegraphics[width=0.45\linewidth]{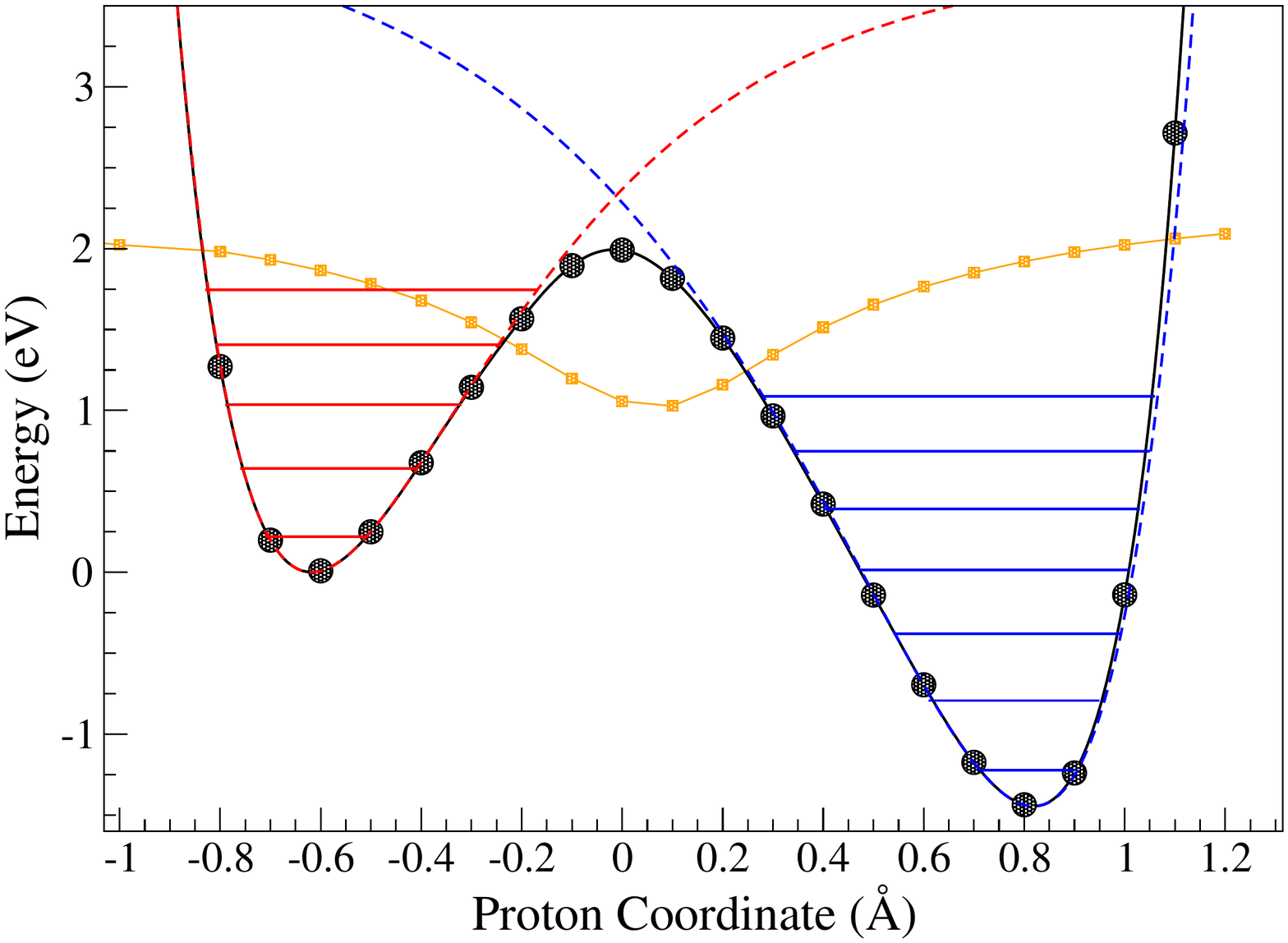} \\
\end{tabular}
\caption{Proton potentials $V(x)$: (a) for the reactant (black circles), the product (green diamonds), the intermediate (red squares), and the transition state (blue triangles) geometries, (b)-(c)-(d) proton potential together with the Morse curves and the $\beta$ HOMO-LUMO gap (orange squares) respectively for the reactant, the intermediate, and the product state geometries. In (c) the $\beta$ HOMO-LUMO gap is divided by 3 for illustration purposes. In all cases, the left well correspond to the hole localized on the TiO$_2$ surface with the proton in the methanol molecule, and the right well  to the hole localized on the adsorbed species and the proton bonded to the bridge oxygen.}
\label{proton_potentials}
\end{figure*}

As specified by Georgievskii and Stuchebrukhova \cite{georgievskii2000concerted}, the proton potential $V(x)$ appearing in Eq. (\ref{wkb}) should be calculated with all other atoms except the transferring proton in a fixed geometry  such that there is a resonance between the vibrational states in both potential wells. As an initial approach, and to keep a reasonable computational workload, $V(x)$ is calculated for the reactant geometry, the product geometry, the transition state geometry, and an intermediate geometry as indicated by $(*)$ in Fig. \ref{NEB-paths}. For these calculations, the positions of all the atoms except the transferring proton are fixed, and the energy is calculated along the direction joining the donor and acceptor oxygen atoms. In contrast to more symmetrical molecular systems such as phenoxyl$/$phenol and benzyl$/$toluene \cite{skone2006calculation}, we do not observe a  double-well potential at the transition state but a rather flat potential, as shown in Fig. \ref{proton_potentials}~(a), suggesting that the proton will be delocalized over both wells. Calculation of the proton vibrational states is then not performed for the transition state geometry.

%%%%%%%%%%%%%%%%%%%%%%%%%%%%%%%%%%%%%%%%%%%%%%%%%%%%%%%%%%%%%%%%%%%%%%%%%%%%%%%%%%%%%%%%%%%%%%%%%%%%%%%%%%%%%%%%%
%%%%%%%%%%%%%%%%%%%%%%%%%%%%%%%%%%%%%%%%%%%%%%%%%%%%%%%%%%%%%%%%%%%%%%%%%%%%%%%%%%%%%%%%%%%%%%%%%%%%%%%%%%%%%%%%%

%%%%%%%%%%%%%%%%%%%%%%%%%%%%%%%%%%%%%%%%%%%%%%%%%%%%%%%%%%%%%%%%%%%%%%%%%%%%%%%%%%%%%%%%%%%%%%%%%%%%%%%%%%%%%%%%%
\begin{table}[t]
\centering
 \begin{tabular}{c | c c c}
 &  $\quad$ $D_e$ (eV) $\quad$ &  $\quad$ $\alpha$ (\AA$^{-1}$) $\quad$  &  $\quad$ $\omega_0$ (cm$^{-1}$) $\quad$  \\
 \hline
 donor well   &  3.143 & 2.505 & 3263 \\
acceptor well &  3.650 & 2.752 & 3863 \\
 \end{tabular}
 \caption{Morse potentials parameters for the reactant geometry as shown in Fig. \ref{proton_potentials}~(b).}
\label{morse-data}
\end{table}
%%%%%%%%%%%%%%%%%%%%%%%%%%%%%%%%%%%%%%%%%%%%%%%%%%%%%%%%%%%%%%%%%%%%%%%%%%%%%%%%%%%%%%%%%%%%%%%%%%%%%%%%%%%%%%%%%

As shown in Fig. \ref{proton_potentials}~(b)-(c)-(d), the two wells of the proton potentials are fitted by Morse potentials, respectively for the reactant state, the intermediate state, and the product state. At the reactant geometry, the vibrational ground state in the donor well is in near resonance with the first excited state in the acceptor well. For the intermediate geometry, the two ground states on both wells are almost in near resonance. Finally, for product geometry the system is not in the ``tunneling resonance regime''. In all cases, we observe a decrease of the vertical energy level of the hole localized on the TiO$_2$ surface when the proton moves until the height of the barrier is passed, and an increase after that where the hole is now localized on the dissociated molecule, as shown by orange squares in Fig. (\ref{proton_potentials}). In the following, vibronic couplings and rate constants are calculated for the reactant geometry. The Morse data for the reactant geometry are shown in Table \ref{morse-data}. Fig. \ref{spin-densities}-(e)-(f) show the spin densities for two structures along the proton coordinate $x$ where we see that the transferring proton carries with itself some $\beta$-density. The corresponding densities of states are shown in Fig. \ref{dos}-(e)-(f).

%%%%%%%%%%%%%%%%%%%%%%%%%%%%%%%%%%%%%%%%%%%%%%%%%%%%%%%%%%%%%%%%%%%%%%%%%%%%%%%%%%%%%%%%%%%%%%%%%%%%%%%%%%%%%%%%%
\begin{center}
\begin{table*}[t]
\centering
\begin{tabular}{ c c c c c c c}
$\quad$ $\Delta G_0$ $\quad$& $\quad$ $\Delta G^{\ddagger}$ $\quad$& $\qquad$ $E_{\text{tun}}$ $\qquad$ & $\qquad$ $|H_{_{\text{DA}}}|$ $\qquad$ & $\qquad$ $\lambda$ $\qquad$& $\quad$ $\langle \varphi_{0} | \varphi_{1} \rangle$ $\quad$\\ 
\hline
 $-$0.34 & 0.60 & 0.199 & 0.186 & 0.091 & 5.8$\times$10$^{-6}$ \\  
\vspace{0.00cm} \\
  $p$  &  $\kappa$ &   $V_{\text{vib}}^{\text{(na)}}$& $V_{\text{vib}}^{\text{(sc)}}$  &  $V_{\text{vib}}^{\text{(ad)}}$ & $\left| E_{\nu}^{(1)}-E_{\mu}^{(0)} \right|$ \\ 
 \hline
  0.045 & 0.452 & 1.1$\times$10$^{-6}$ & 1.5$\times$10$^{-6}$ &3.3$\times$10$^{-6}$ & 0.016 \\
\vspace{0.00cm} \\
$\tau_e$ & $\tau_p$  &    $k^{\text{(na)}}_{_{\text{PCET}}}$  &  $k^{\text{(sc)}}_{_{\text{PCET}}}$  & $k^{\text{(ad)}}_{_{\text{PCET}}}$  &  $k_{_{\text{TST}}}$ &  \\ 
 \hline
3.54 & 0.16  &   4.7$\times$10$^{1}$ & 8.8$\times$10$^{1}$ & 4.3$\times$10$^{2}$ & 5.2$\times$10$^{2}$   \\ 
\end{tabular}
\caption{Calculated parameters for the calculation of the rate constants. Calculations are done for the reactant geometry (see Fig. \ref{proton_potentials}.(b)). Vibronic coupling are given for $\mu=0$ and $\nu=1$, which correspond to the near resonant states in the reactant geometry and which largely dominate the rate constants. All energies are in eV, time in fs, and rate constants in s$^{-1}$. $k_{_{\text{PCET}}}$ are calculated from Eq. (\ref{k_me}) and $k_{_{\text{TST}}}$ from Eq. (\ref{k_tst}). See the main text for a description of the different parameters.}
\label{results}
\end{table*}
\end{center}
%%%%%%%%%%%%%%%%%%%%%%%%%%%%%%%%%%%%%%%%%%%%%%%%%%%%%%%%%%%%%%%%%%%%%%%%%%%%%%%%%%%%%%%%%%%%%%%%%%%%%%%%%%%%%%%%%

Using Eq. (\ref{vib_sc}), the vibronic coupling is calculated to be 1.5$\times$10$^{-6}$~eV, indicating a vibronically nonadiabatic reaction and justifying the use of the theory presented in section II.1. The adiabaticity parameters are calculated to be $p = 0.045$ and $\kappa = 0.452$, indicating an intermediate regime between electronically adiabatic and nonadiabatic regimes. We now have calculated all the required quantities to calculate the rate constant from Eq. (\ref{k_me}), which is calculated to be 8.8$\times$10$^{1}$~s$^{-1}$. Interestingly, this rate is not very far from the TST rate calculated to be 5.2$\times$10$^{2}$~s$^{-1}$. Finally, the different limit rates are in a small range of values in the order:
\begin{equation}\label{order}
k^{\text{(na)}}_{_{\text{PCET}}} < k^{\text{(sc)}}_{_{\text{PCET}}} < k^{\text{(ad)}}_{_{\text{PCET}}} < k_{_{\text{TST}}},
\end{equation}
showing a coherent ordering of the PCET rate constants across the different regimes of vibronic and electronic nonadiabaticity. The principal numerical results are gathered in table~\ref{results}. Of course, these results are relevant only for the particular path we have chosen, which correspond to a particular hole localization site, a particular adsorption geometry, and a particular bridge oxygen receiving the transferring proton. Also, a sampling of the different possible resonante structures should be calculated, for instance from \textit{ab initio} molecular dynamics.

\section{V. Conclusion}

In this work, we have studied the kinetics of the first step of methanol photo-dissociation on rutile TiO$_2$ surface using and slightly modifying the Stuchebrukhov-Hammes-Schiffer theory \cite{hammes2008proton,hammes2010theory}, together with the Georgievskii-Stuchebrukhova theory to calculate the vibronic couplings \cite{georgievskii2000concerted}. In this framework, we have studied the OH bond cleavage of adsorbed methanol for a given hole trapping site, a given adsorption geometry, and a given reaction path. the reaction is found to be vibronically nonadiabatic and in an intermediate regime between electronically adiabatic and electronically nonadiabatic. In particular, we have proposed a modified PCET rate constant equation to enforce the near-resonance condition on the proton vibrational wave functions given by Eq. (\ref{k_me}) allowing us to obtain a coherent ordering of the PCET rate constants across the different regimes of vibronic and electronic nonadiabaticity (see Eq. (\ref{order})). We have presented here the first step of our methodology development in computational photocatalysis. We believe that these results are general and can be applied to a large range of photocatalytic reactions on oxide semiconductor surfaces.

\begin{acknowledgments}
\section{Acknowledgement}
We are grateful for funding support by the National Natural Science Foundation of China (Grant Nos. 21991151, 91745103 and 21861132015).
\end{acknowledgments}

%\appendix

\section{Appendices}

\subsection{A. Electron-Proton Vibronic States}\label{Morse}

To calculate the quantities $P_{\mu}$, $E_{\mu\nu}$ and $S_{\mu\nu} = \langle \varphi_{\mu}|\varphi_{\nu} \rangle$, where $\varphi_{\mu/\nu}$ are the proton vibrational wavefunctions, the proton potential wells are approximated by Morse potentials of the form:
\begin{equation}\label{Morse-pot}
V(r) = D_e \left( 1 - e^{-\alpha (r - r_{\text{eq}})} \right)^2, 
\end{equation}
where $r$ is the distance between the transferring proton and the nearby atom to which it is bound, $r_{\text{eq}}$ the equilibrium bond distance, $D_e$ the dissociation energy, and $\alpha = \sqrt{k_e/2D_e}$, where $k_e$ is the force constant at the minimum of the well.

Analytical solutions of the one-dimensional Schr\"odinger equation are available for the eigenfunctions and eigenvalues of the Morse potential \cite{dahl1988morse}. The eigenfunctions are given by:
\begin{equation}\label{proton_WF}
\varphi_n(r) = N_n \xi(r)^{\beta-n-1/2} e^{-\xi(r)/2} L_n^{(2\beta-2n-1)}(\xi),
\end{equation}
where:
\begin{equation}\label{beta-xi}
\beta = \frac{\sqrt{2mD_e}}{\alpha \hbar}, \quad \text{and} \quad
\xi(r) = 2 \beta e^{-\alpha(r-r_{\text{eq}})},
\end{equation}
where the normalization constant $N_n$ has been calculated numerically. Finally, $L_n^{(2\beta-2n-1)}(\xi)$ are the generalized Laguerre polynomials. It is known that analytic expression for the Morse wave functions is computationally unstable because of the summation of an alternating series in evaluating Laguerre functions, especially for higher excited states. Dahl and Springborg have proposed a recurrence relation which allows to circumvent this problem, given by \cite{dahl1988morse}:
\begin{eqnarray}
L_n^{(\sigma)}(z) &=& \frac{1}{n} \bigg[ \left( 2n-1+\sigma - z \right) L_{n-1}^{(\sigma)}(z) \nonumber \\
&-& \left( n-1+\sigma \right) L_{n-2}^{(\sigma)}(z) \bigg],
\end{eqnarray} 
with the two first polynomials given by:
\begin{equation}
L_0^{(\sigma)}(z) = 1 \quad \text{and} \quad L_1^{(\sigma)}(z) = 1 + \sigma - z.
\end{equation}

The number of bound states is $\lfloor{\beta + 1/2}\rfloor$, where $\lfloor x \rfloor$ denotes the largest integer smaller than $x$. The corresponding energies are given by:
\begin{equation}\label{proton_energy}
E_n = \left[ \left( n+\frac{1}{2}\right) - \frac{1}{2\beta} \left( n+\frac{1}{2}\right)^2 \right] \hbar \omega_0,
\end{equation}
where 
\begin{equation}\label{omega}
\omega_0 = \left( \frac{2 D_e \alpha^2}{m} \right)^{1/2}.
\end{equation}
From (\ref{proton_energy}), we can calculate the Boltzmann population at a given temperature from:
\begin{equation}
P_{\mu} = \frac{e^{- E_{\mu} / k_{_{\text{B}}} T }}{\sum_{\mu} e^{- E_{\mu} / k_{_{\text{B}}} T}}.
\end{equation}

\subsection{B. Electronic Coupling}\label{elec_coupling}

Several methods exist which allow to calculate the electronic coupling $H_{_{\text{DA}}}$ involved in Eq. (\ref{vib_na}) \cite{blumberger2015recent}. Here, we have implemented the method developed by Kondov \textit{et al.} \cite{kondov2007quantum} and referred to as projector-operator diabatization (POD) method \cite{futera2017electronic}. The POD method is a post-processing diabatization method whose starting point is the Kohn-Sham (KS) matrix obtained from a converged \textit{ab initio} calculation $H_{_{\text{KS}}}$. The adiabatic electronic states $| \Psi_i \rangle$ obtained from standard DFT calculation are expressed as a linear combination of atomic-orbital basis set: $| \Psi_i \rangle = \sum_{j} c_{ij}| \phi_j \rangle$. For plane-wave codes, it should be possible to perform a projection on localized orbitals such as Wannier functions. The set of atomic orbitals of the overall system, $| \phi_j \rangle$, is divided into two groups, namely, the donor group ($| \phi_j^d \rangle$), which comprises the orbitals centered at the atoms of the donor, and the acceptor group ($| \phi_j^a \rangle$), which includes the orbitals centered at the acceptor. In our case, atoms belonging to TiO$_2$ are considered as being part of the hole donor (together with the water molecules and the bottom methanol molecule) and atoms belonging to active methanol molecule are considered as being part of the hole acceptor. The transferring proton can be either chosen as being part of the donor or the acceptor.

Localized basis functions, such as Gaussian-type orbitals we are using here, are in general not orthogonalized, and the Hamiltonian is expressed in a new orthogonalized basis set according to the L\"owdin procedure~\cite{lowdin1950non,mayer2002lowdin}:
\begin{equation}
\tilde{H}_{_{\text{KS}}} = S^{-1/2} H_{_{\text{KS}}} S^{-1/2},
\end{equation}
where $S$ denotes the electronic orbital overlap matrix with elements $S_{kl} = \langle \phi_k  | \phi_l \rangle$, which can be for instance directly printed out from a CP2K calculation. To obtain $S^{-1/2}$, we first diagonalize the overlap matrix using a unitary matrix $U$ (i.e. $U^{\dagger}U=UU^{\dagger}=1$): $S_{\text{diag}} = U^{\dagger}S U$. $S$~being a real matrix, $U$ is orthogonal (i.e. $U^{\text{T}}U=UU^{\text{T}}=1$). The eigenvalues of $S$ are always positive, which allows to replace the diagonal elements of $S_{\text{diag}}$ by their square roots, giving the matrix $S_{\text{diag}}^{1/2}$. After calculating $S^{1/2} = U S_{\text{diag}}^{1/2}U^{\dagger}$, we obtain $S^{-1/2} = (S^{1/2})^{-1} = U S_{\text{diag}}^{-1/2}U^{\dagger} $ (See appendix J of Ref. \cite{piela2013ideas}). Direct diagonalization of $\tilde{H}_{_{\text{KS}}}$  gives the eigenvalues obtain from the initial \textit{ab initio} calculation.

The matrix $\tilde{H}_{_{\text{KS}}}$ can be arranged in the following donor-acceptor block structure according to the previously chosen separation:
\begin{equation}
\tilde{H}_{_{\text{KS}}} =
\left(\begin{array}{cc} \tilde{H} _{dd}& \tilde{H}_{da} \\ 
\tilde{H}_{ad} & \tilde{H}_{aa} \end{array}\right).
\end{equation}
Separate diagonalization of the two (donor and acceptor) blocks of the KS matrix $\tilde{H}_{\alpha \alpha}$ via $
\bar{H}_{\alpha \alpha} =  U^{\dagger}_{\alpha}  \tilde{H}_{\alpha \alpha} U_{\alpha}$, 
and transformation of the off-diagonal parts with the corresponding eigenstates in the two blocks via
$\bar{H}_{\alpha \beta} =  U^{\dagger}_{\alpha}  \tilde{H}_{\alpha \beta} U_{\beta}$ (where $\alpha$ and $\beta$ denote either the donor (d) or the acceptor (a)), 
result in the following block structure form:
\begin{eqnarray}\label{H-coupling}
\bar{H}_{_{\text{KS}}} &=&
\left(\begin{array}{cc} \bar{H} _{dd}& \bar{H}_{da} \\ 
\bar{H}_{ad} & \bar{H}_{aa} \end{array}\right) \\
\quad \nonumber \\
&=& \left(\begin{array}{cccccc} \varepsilon_{d,1} & 0 & \cdots & & & \\
0 & \varepsilon_{d,2} & \cdots & & \bar{H}_{da}  & \\ 
\vdots & \vdots & & & &  \\
 &  &  & \varepsilon_{a,1} & 0 & \cdots \\
 & \bar{H}_{ad} & & 0 & \varepsilon_{a,2} & \cdots \\
 &  &  & \vdots & \vdots & \end{array} \right), \nonumber
\end{eqnarray}
where $\varepsilon_{\alpha,i}$ are the one-electron energies of the diabatic states of the donor and acceptor, and $\bar{H}_{\alpha\beta}$ are the coupling between them. In our case, $\bar{H}_{ad} = \bar{H}_{da}$. We note here that by using CP2K at the hybrid level in the ADMM approach when calculating the electronic couplings, it is required to turn off the purification procedure in order to obtain the correct Kohn-Sham matrix elements~\cite{guidon2010auxiliary}.

The corresponding donor and acceptor molecular orbitals, $| \bar{\phi}_n^{\alpha} \rangle$, are given as the eigenvectors of $\bar{H}_{\alpha \alpha}$ and are related to the orthogonalized atomic orbitals $| \tilde{\phi}_j^{\alpha} \rangle$ and the original atomic orbitals $| \phi_l \rangle$ via \cite{kondov2007quantum}:
\begin{eqnarray}\label{MO_part}
| \bar{\phi}_n^{\alpha} \rangle &=& \sum_j \left( U_{\alpha} \right)_{jn} | \tilde{\phi}_j^{\alpha} \rangle \nonumber \\
&=& \sum_{js} \left( U_{\alpha} \right)_{jn} \left( S^{-1/2} \right)_{sj} | \phi_s \rangle.
\end{eqnarray}
Identifying the donor (or acceptor) state with one of the states $| \bar{\phi}_n^{d} \rangle$ and the acceptor (or donor) states with one of the state $| \bar{\phi}_m^{a} \rangle$, the electronic coupling is given by $ | H_{_{\text{DA}}} | = |\bar{H}_{da,nm}|$. After separation of the system into a donor and an acceptor part, the molecular orbitals of interest can be mixed. We have opted for an identification by visualization in which the molecular orbitals after separation, given by equation (\ref{MO_part}), are compared either to the molecular orbitals of the full system before separation or to the molecular orbitals of isolated parts.

\subsection{C. Reorganization Energy}
As proposed in Ref. \cite{ghosh2017theoretical,fernandez2012theoretical}, the equilibrium energies of the isolated fragments are combined to compute the inner-sphere reorganization energy with a variant of the four-point model. Considering the methanol dissociation where a proton $\text{H}^+$ and a hole $h^+$ are exchanged between the molecule and the semiconductor:
\begin{equation}
\text{TiO}_2/h^+ + \text{CH}_3\text{OH} \longrightarrow \text{TiO}_2/\text{H}^+ + \text{CH}_3\text{O}^{\Bigcdot},
\end{equation}
the inner-sphere reorganization energy can be calculated as follows :
\begin{eqnarray}\label{lambda_in}
\lambda = \frac{1}{2} 
&\big[& E_{\text{TiO}_2/h^+ }(\text{TiO}_2/\text{H}^+) + E_{\text{CH}_3\text{OH}}(\text{CH}_3\text{O}^{\Bigcdot})\nonumber \\
&-& E_{\text{TiO}_2/h^+ }(\text{TiO}_2/h^+  ) - E_{\text{CH}_3\text{OH}}(\text{CH}_3\text{OH}) \big] \nonumber \\
+ \frac{1}{2} &\big[& E_{\text{TiO}_2/\text{H}^+}(\text{TiO}_2/h^+ ) + E_{\text{CH}_3\text{O}^{\Bigcdot}}(\text{CH}_3\text{OH}) \nonumber \\
&-& E_{\text{TiO}_2/\text{H}^+}(\text{TiO}_2/\text{H}^+) - E_{\text{CH}_3\text{O}^{\Bigcdot}}(\text{CH}_3\text{O}^{\Bigcdot}) \big], \nonumber \\
\quad
\end{eqnarray}
where $E_{\text{A}}(\text{B})$ is the energy of state A at the optimized geometry of state B. These energies correspond to equilibrium energy of the particular fragment when A and B are the same, and correspond to a nonequilibrium geometry otherwise. If B is protonated and not A, the proton on B is removed, whereas when B is not protonated and A is, the proton position is optimized while keeping the other atoms at fixed positions.

\bibliography{article}

\begin{thebibliography}{62}%
\makeatletter
\providecommand \@ifxundefined [1]{%
 \@ifx{#1\undefined}
}%
\providecommand \@ifnum [1]{%
 \ifnum #1\expandafter \@firstoftwo
 \else \expandafter \@secondoftwo
 \fi
}%
\providecommand \@ifx [1]{%
 \ifx #1\expandafter \@firstoftwo
 \else \expandafter \@secondoftwo
 \fi
}%
\providecommand \natexlab [1]{#1}%
\providecommand \enquote  [1]{``#1''}%
\providecommand \bibnamefont  [1]{#1}%
\providecommand \bibfnamefont [1]{#1}%
\providecommand \citenamefont [1]{#1}%
\providecommand \href@noop [0]{\@secondoftwo}%
\providecommand \href [0]{\begingroup \@sanitize@url \@href}%
\providecommand \@href[1]{\@@startlink{#1}\@@href}%
\providecommand \@@href[1]{\endgroup#1\@@endlink}%
\providecommand \@sanitize@url [0]{\catcode `\\12\catcode `\$12\catcode
  `\&12\catcode `\#12\catcode `\^12\catcode `\_12\catcode `\%12\relax}%
\providecommand \@@startlink[1]{}%
\providecommand \@@endlink[0]{}%
\providecommand \url  [0]{\begingroup\@sanitize@url \@url }%
\providecommand \@url [1]{\endgroup\@href {#1}{\urlprefix }}%
\providecommand \urlprefix  [0]{URL }%
\providecommand \Eprint [0]{\href }%
\providecommand \doibase [0]{http://dx.doi.org/}%
\providecommand \selectlanguage [0]{\@gobble}%
\providecommand \bibinfo  [0]{\@secondoftwo}%
\providecommand \bibfield  [0]{\@secondoftwo}%
\providecommand \translation [1]{[#1]}%
\providecommand \BibitemOpen [0]{}%
\providecommand \bibitemStop [0]{}%
\providecommand \bibitemNoStop [0]{.\EOS\space}%
\providecommand \EOS [0]{\spacefactor3000\relax}%
\providecommand \BibitemShut  [1]{\csname bibitem#1\endcsname}%
\let\auto@bib@innerbib\@empty
%</preamble>
\bibitem [{\citenamefont {van~de Krol}\ and\ \citenamefont
  {Parkinson}(2017)}]{van2017perspectives}%
  \BibitemOpen
  \bibfield  {author} {\bibinfo {author} {\bibfnamefont {R.}~\bibnamefont
  {van~de Krol}}\ and\ \bibinfo {author} {\bibfnamefont {B.~A.}\ \bibnamefont
  {Parkinson}},\ }\href@noop {} {\bibfield  {journal} {\bibinfo  {journal} {MRS
  Energy \& Sustainability}\ }\textbf {\bibinfo {volume} {4}},\ \bibinfo
  {pages} {E13} (\bibinfo {year} {2017})}\BibitemShut {NoStop}%
\bibitem [{\citenamefont {Wang}\ \emph {et~al.}(2019)\citenamefont {Wang},
  \citenamefont {Li},\ and\ \citenamefont {Domen}}]{wang2019recent}%
  \BibitemOpen
  \bibfield  {author} {\bibinfo {author} {\bibfnamefont {Z.}~\bibnamefont
  {Wang}}, \bibinfo {author} {\bibfnamefont {C.}~\bibnamefont {Li}}, \ and\
  \bibinfo {author} {\bibfnamefont {K.}~\bibnamefont {Domen}},\ }\href@noop {}
  {\bibfield  {journal} {\bibinfo  {journal} {Chemical Society Reviews}\
  }\textbf {\bibinfo {volume} {48}},\ \bibinfo {pages} {2109} (\bibinfo {year}
  {2019})}\BibitemShut {NoStop}%
\bibitem [{\citenamefont {Hisatomi}\ and\ \citenamefont
  {Domen}(2019)}]{hisatomi2019reaction}%
  \BibitemOpen
  \bibfield  {author} {\bibinfo {author} {\bibfnamefont {T.}~\bibnamefont
  {Hisatomi}}\ and\ \bibinfo {author} {\bibfnamefont {K.}~\bibnamefont
  {Domen}},\ }\href@noop {} {\bibfield  {journal} {\bibinfo  {journal} {Nature
  Catalysis}\ }\textbf {\bibinfo {volume} {2}},\ \bibinfo {pages} {387}
  (\bibinfo {year} {2019})}\BibitemShut {NoStop}%
\bibitem [{\citenamefont {Fujishima}\ and\ \citenamefont
  {Honda}(1972)}]{fujishima1972electrochemical}%
  \BibitemOpen
  \bibfield  {author} {\bibinfo {author} {\bibfnamefont {A.}~\bibnamefont
  {Fujishima}}\ and\ \bibinfo {author} {\bibfnamefont {K.}~\bibnamefont
  {Honda}},\ }\href@noop {} {\bibfield  {journal} {\bibinfo  {journal}
  {nature}\ }\textbf {\bibinfo {volume} {238}},\ \bibinfo {pages} {37}
  (\bibinfo {year} {1972})}\BibitemShut {NoStop}%
\bibitem [{\citenamefont {Thompson}\ and\ \citenamefont
  {Yates}(2006)}]{thompson2006surface}%
  \BibitemOpen
  \bibfield  {author} {\bibinfo {author} {\bibfnamefont {T.~L.}\ \bibnamefont
  {Thompson}}\ and\ \bibinfo {author} {\bibfnamefont {J.~T.}\ \bibnamefont
  {Yates}},\ }\href@noop {} {\bibfield  {journal} {\bibinfo  {journal}
  {Chemical Reviews}\ }\textbf {\bibinfo {volume} {106}},\ \bibinfo {pages}
  {4428} (\bibinfo {year} {2006})}\BibitemShut {NoStop}%
\bibitem [{\citenamefont {Fujishima}\ \emph {et~al.}(2008)\citenamefont
  {Fujishima}, \citenamefont {Zhang},\ and\ \citenamefont
  {Tryk}}]{FUJISHIMA2008515}%
  \BibitemOpen
  \bibfield  {author} {\bibinfo {author} {\bibfnamefont {A.}~\bibnamefont
  {Fujishima}}, \bibinfo {author} {\bibfnamefont {X.}~\bibnamefont {Zhang}}, \
  and\ \bibinfo {author} {\bibfnamefont {D.~A.}\ \bibnamefont {Tryk}},\ }\href
  {\doibase https://doi.org/10.1016/j.surfrep.2008.10.001} {\bibfield
  {journal} {\bibinfo  {journal} {Surface Science Reports}\ }\textbf {\bibinfo
  {volume} {63}},\ \bibinfo {pages} {515 } (\bibinfo {year}
  {2008})}\BibitemShut {NoStop}%
\bibitem [{\citenamefont {Henderson}(2011)}]{henderson2011surface}%
  \BibitemOpen
  \bibfield  {author} {\bibinfo {author} {\bibfnamefont {M.~A.}\ \bibnamefont
  {Henderson}},\ }\href@noop {} {\bibfield  {journal} {\bibinfo  {journal}
  {Surface Science Reports}\ }\textbf {\bibinfo {volume} {66}},\ \bibinfo
  {pages} {185} (\bibinfo {year} {2011})}\BibitemShut {NoStop}%
\bibitem [{\citenamefont {Guo}\ \emph {et~al.}(2016)\citenamefont {Guo},
  \citenamefont {Zhou}, \citenamefont {Ma}, \citenamefont {Ren}, \citenamefont
  {Fan},\ and\ \citenamefont {Yang}}]{guo2016elementary}%
  \BibitemOpen
  \bibfield  {author} {\bibinfo {author} {\bibfnamefont {Q.}~\bibnamefont
  {Guo}}, \bibinfo {author} {\bibfnamefont {C.}~\bibnamefont {Zhou}}, \bibinfo
  {author} {\bibfnamefont {Z.}~\bibnamefont {Ma}}, \bibinfo {author}
  {\bibfnamefont {Z.}~\bibnamefont {Ren}}, \bibinfo {author} {\bibfnamefont
  {H.}~\bibnamefont {Fan}}, \ and\ \bibinfo {author} {\bibfnamefont
  {X.}~\bibnamefont {Yang}},\ }\href@noop {} {\bibfield  {journal} {\bibinfo
  {journal} {Chemical Society Reviews}\ }\textbf {\bibinfo {volume} {45}},\
  \bibinfo {pages} {3701} (\bibinfo {year} {2016})}\BibitemShut {NoStop}%
\bibitem [{\citenamefont {Kawai}\ and\ \citenamefont
  {Sakata}(1980)}]{kawai1980photocatalytic}%
  \BibitemOpen
  \bibfield  {author} {\bibinfo {author} {\bibfnamefont {T.}~\bibnamefont
  {Kawai}}\ and\ \bibinfo {author} {\bibfnamefont {T.}~\bibnamefont {Sakata}},\
  }\href@noop {} {\bibfield  {journal} {\bibinfo  {journal} {Journal of the
  Chemical Society, Chemical Communications}\ ,\ \bibinfo {pages} {694}}
  (\bibinfo {year} {1980})}\BibitemShut {NoStop}%
\bibitem [{\citenamefont {Wei}\ \emph {et~al.}(2015)\citenamefont {Wei},
  \citenamefont {Jin}, \citenamefont {Huang}, \citenamefont {Dai},
  \citenamefont {Ma}, \citenamefont {Li},\ and\ \citenamefont
  {Yang}}]{wei2015direct}%
  \BibitemOpen
  \bibfield  {author} {\bibinfo {author} {\bibfnamefont {D.}~\bibnamefont
  {Wei}}, \bibinfo {author} {\bibfnamefont {X.}~\bibnamefont {Jin}}, \bibinfo
  {author} {\bibfnamefont {C.}~\bibnamefont {Huang}}, \bibinfo {author}
  {\bibfnamefont {D.}~\bibnamefont {Dai}}, \bibinfo {author} {\bibfnamefont
  {Z.}~\bibnamefont {Ma}}, \bibinfo {author} {\bibfnamefont {W.-X.}\
  \bibnamefont {Li}}, \ and\ \bibinfo {author} {\bibfnamefont {X.}~\bibnamefont
  {Yang}},\ }\href@noop {} {\bibfield  {journal} {\bibinfo  {journal} {The
  Journal of Physical Chemistry C}\ }\textbf {\bibinfo {volume} {119}},\
  \bibinfo {pages} {17748} (\bibinfo {year} {2015})}\BibitemShut {NoStop}%
\bibitem [{\citenamefont {Feng}\ \emph {et~al.}(2016)\citenamefont {Feng},
  \citenamefont {Tan}, \citenamefont {Tang}, \citenamefont {Zheng},
  \citenamefont {Shi}, \citenamefont {Cui}, \citenamefont {Shao}, \citenamefont
  {Zhao}, \citenamefont {Zhao},\ and\ \citenamefont
  {Wang}}]{feng2016temperature}%
  \BibitemOpen
  \bibfield  {author} {\bibinfo {author} {\bibfnamefont {H.}~\bibnamefont
  {Feng}}, \bibinfo {author} {\bibfnamefont {S.}~\bibnamefont {Tan}}, \bibinfo
  {author} {\bibfnamefont {H.}~\bibnamefont {Tang}}, \bibinfo {author}
  {\bibfnamefont {Q.}~\bibnamefont {Zheng}}, \bibinfo {author} {\bibfnamefont
  {Y.}~\bibnamefont {Shi}}, \bibinfo {author} {\bibfnamefont {X.}~\bibnamefont
  {Cui}}, \bibinfo {author} {\bibfnamefont {X.}~\bibnamefont {Shao}}, \bibinfo
  {author} {\bibfnamefont {A.}~\bibnamefont {Zhao}}, \bibinfo {author}
  {\bibfnamefont {J.}~\bibnamefont {Zhao}}, \ and\ \bibinfo {author}
  {\bibfnamefont {B.}~\bibnamefont {Wang}},\ }\href@noop {} {\bibfield
  {journal} {\bibinfo  {journal} {The Journal of Physical Chemistry C}\
  }\textbf {\bibinfo {volume} {120}},\ \bibinfo {pages} {5503} (\bibinfo {year}
  {2016})}\BibitemShut {NoStop}%
\bibitem [{\citenamefont {Guo}\ \emph {et~al.}(2012)\citenamefont {Guo},
  \citenamefont {Xu}, \citenamefont {Ren}, \citenamefont {Yang}, \citenamefont
  {Ma}, \citenamefont {Dai}, \citenamefont {Fan}, \citenamefont {Minton},\ and\
  \citenamefont {Yang}}]{guo2012stepwise}%
  \BibitemOpen
  \bibfield  {author} {\bibinfo {author} {\bibfnamefont {Q.}~\bibnamefont
  {Guo}}, \bibinfo {author} {\bibfnamefont {C.}~\bibnamefont {Xu}}, \bibinfo
  {author} {\bibfnamefont {Z.}~\bibnamefont {Ren}}, \bibinfo {author}
  {\bibfnamefont {W.}~\bibnamefont {Yang}}, \bibinfo {author} {\bibfnamefont
  {Z.}~\bibnamefont {Ma}}, \bibinfo {author} {\bibfnamefont {D.}~\bibnamefont
  {Dai}}, \bibinfo {author} {\bibfnamefont {H.}~\bibnamefont {Fan}}, \bibinfo
  {author} {\bibfnamefont {T.~K.}\ \bibnamefont {Minton}}, \ and\ \bibinfo
  {author} {\bibfnamefont {X.}~\bibnamefont {Yang}},\ }\href@noop {} {\bibfield
   {journal} {\bibinfo  {journal} {Journal of the American Chemical Society}\
  }\textbf {\bibinfo {volume} {134}},\ \bibinfo {pages} {13366} (\bibinfo
  {year} {2012})}\BibitemShut {NoStop}%
\bibitem [{\citenamefont {Shen}\ and\ \citenamefont
  {Henderson}(2011)}]{shen2011identification}%
  \BibitemOpen
  \bibfield  {author} {\bibinfo {author} {\bibfnamefont {M.}~\bibnamefont
  {Shen}}\ and\ \bibinfo {author} {\bibfnamefont {M.~A.}\ \bibnamefont
  {Henderson}},\ }\href@noop {} {\bibfield  {journal} {\bibinfo  {journal} {The
  Journal of Physical Chemistry Letters}\ }\textbf {\bibinfo {volume} {2}},\
  \bibinfo {pages} {2707} (\bibinfo {year} {2011})}\BibitemShut {NoStop}%
\bibitem [{\citenamefont {Zhang}\ \emph {et~al.}(2017)\citenamefont {Zhang},
  \citenamefont {Peng}, \citenamefont {Wang},\ and\ \citenamefont
  {Hu}}]{zhang2017identifying}%
  \BibitemOpen
  \bibfield  {author} {\bibinfo {author} {\bibfnamefont {J.}~\bibnamefont
  {Zhang}}, \bibinfo {author} {\bibfnamefont {C.}~\bibnamefont {Peng}},
  \bibinfo {author} {\bibfnamefont {H.}~\bibnamefont {Wang}}, \ and\ \bibinfo
  {author} {\bibfnamefont {P.}~\bibnamefont {Hu}},\ }\href@noop {} {\bibfield
  {journal} {\bibinfo  {journal} {ACS Catalysis}\ }\textbf {\bibinfo {volume}
  {7}},\ \bibinfo {pages} {2374} (\bibinfo {year} {2017})}\BibitemShut
  {NoStop}%
\bibitem [{\citenamefont {Chu}\ \emph {et~al.}(2016)\citenamefont {Chu},
  \citenamefont {Saidi}, \citenamefont {Zheng}, \citenamefont {Xie},
  \citenamefont {Lan}, \citenamefont {Prezhdo}, \citenamefont {Petek},\ and\
  \citenamefont {Zhao}}]{chu2016ultrafast}%
  \BibitemOpen
  \bibfield  {author} {\bibinfo {author} {\bibfnamefont {W.}~\bibnamefont
  {Chu}}, \bibinfo {author} {\bibfnamefont {W.~A.}\ \bibnamefont {Saidi}},
  \bibinfo {author} {\bibfnamefont {Q.}~\bibnamefont {Zheng}}, \bibinfo
  {author} {\bibfnamefont {Y.}~\bibnamefont {Xie}}, \bibinfo {author}
  {\bibfnamefont {Z.}~\bibnamefont {Lan}}, \bibinfo {author} {\bibfnamefont
  {O.~V.}\ \bibnamefont {Prezhdo}}, \bibinfo {author} {\bibfnamefont
  {H.}~\bibnamefont {Petek}}, \ and\ \bibinfo {author} {\bibfnamefont
  {J.}~\bibnamefont {Zhao}},\ }\href@noop {} {\bibfield  {journal} {\bibinfo
  {journal} {Journal of the American Chemical Society}\ }\textbf {\bibinfo
  {volume} {138}},\ \bibinfo {pages} {13740} (\bibinfo {year}
  {2016})}\BibitemShut {NoStop}%
\bibitem [{\citenamefont {Migani}\ and\ \citenamefont
  {Blancafort}(2016)}]{migani2016excitonic}%
  \BibitemOpen
  \bibfield  {author} {\bibinfo {author} {\bibfnamefont {A.}~\bibnamefont
  {Migani}}\ and\ \bibinfo {author} {\bibfnamefont {L.}~\bibnamefont
  {Blancafort}},\ }\href@noop {} {\bibfield  {journal} {\bibinfo  {journal}
  {Journal of the American Chemical Society}\ }\textbf {\bibinfo {volume}
  {138}},\ \bibinfo {pages} {16165} (\bibinfo {year} {2016})}\BibitemShut
  {NoStop}%
\bibitem [{\citenamefont {Zhu}\ \emph {et~al.}(2017)\citenamefont {Zhu},
  \citenamefont {Duan}, \citenamefont {Ji}, \citenamefont {Song}, \citenamefont
  {Chen}, \citenamefont {Ma},\ and\ \citenamefont {Zhao}}]{zhu2017interfacial}%
  \BibitemOpen
  \bibfield  {author} {\bibinfo {author} {\bibfnamefont {Q.}~\bibnamefont
  {Zhu}}, \bibinfo {author} {\bibfnamefont {R.}~\bibnamefont {Duan}}, \bibinfo
  {author} {\bibfnamefont {H.}~\bibnamefont {Ji}}, \bibinfo {author}
  {\bibfnamefont {W.}~\bibnamefont {Song}}, \bibinfo {author} {\bibfnamefont
  {C.}~\bibnamefont {Chen}}, \bibinfo {author} {\bibfnamefont {W.}~\bibnamefont
  {Ma}}, \ and\ \bibinfo {author} {\bibfnamefont {J.}~\bibnamefont {Zhao}},\
  }\href@noop {} {\bibfield  {journal} {\bibinfo  {journal} {Research on
  Chemical Intermediates}\ }\textbf {\bibinfo {volume} {43}},\ \bibinfo {pages}
  {4997} (\bibinfo {year} {2017})}\BibitemShut {NoStop}%
\bibitem [{\citenamefont {Chen}\ \emph {et~al.}(2015)\citenamefont {Chen},
  \citenamefont {Shi}, \citenamefont {Chang},\ and\ \citenamefont
  {Zhao}}]{chen2015essential}%
  \BibitemOpen
  \bibfield  {author} {\bibinfo {author} {\bibfnamefont {C.}~\bibnamefont
  {Chen}}, \bibinfo {author} {\bibfnamefont {T.}~\bibnamefont {Shi}}, \bibinfo
  {author} {\bibfnamefont {W.}~\bibnamefont {Chang}}, \ and\ \bibinfo {author}
  {\bibfnamefont {J.}~\bibnamefont {Zhao}},\ }\href@noop {} {\bibfield
  {journal} {\bibinfo  {journal} {ChemCatChem}\ }\textbf {\bibinfo {volume}
  {7}},\ \bibinfo {pages} {724} (\bibinfo {year} {2015})}\BibitemShut {NoStop}%
\bibitem [{\citenamefont {Schrauben}\ \emph {et~al.}(2012)\citenamefont
  {Schrauben}, \citenamefont {Hayoun}, \citenamefont {Valdez}, \citenamefont
  {Braten}, \citenamefont {Fridley},\ and\ \citenamefont
  {Mayer}}]{schrauben2012titanium}%
  \BibitemOpen
  \bibfield  {author} {\bibinfo {author} {\bibfnamefont {J.~N.}\ \bibnamefont
  {Schrauben}}, \bibinfo {author} {\bibfnamefont {R.}~\bibnamefont {Hayoun}},
  \bibinfo {author} {\bibfnamefont {C.~N.}\ \bibnamefont {Valdez}}, \bibinfo
  {author} {\bibfnamefont {M.}~\bibnamefont {Braten}}, \bibinfo {author}
  {\bibfnamefont {L.}~\bibnamefont {Fridley}}, \ and\ \bibinfo {author}
  {\bibfnamefont {J.~M.}\ \bibnamefont {Mayer}},\ }\href@noop {} {\bibfield
  {journal} {\bibinfo  {journal} {Science}\ }\textbf {\bibinfo {volume}
  {336}},\ \bibinfo {pages} {1298} (\bibinfo {year} {2012})}\BibitemShut
  {NoStop}%
\bibitem [{\citenamefont {Hoffmann}(2017)}]{hoffmann2017proton}%
  \BibitemOpen
  \bibfield  {author} {\bibinfo {author} {\bibfnamefont {N.}~\bibnamefont
  {Hoffmann}},\ }\href@noop {} {\bibfield  {journal} {\bibinfo  {journal}
  {European Journal of Organic Chemistry}\ }\textbf {\bibinfo {volume}
  {2017}},\ \bibinfo {pages} {1982} (\bibinfo {year} {2017})}\BibitemShut
  {NoStop}%
\bibitem [{\citenamefont {Chen}\ \emph {et~al.}(2013)\citenamefont {Chen},
  \citenamefont {Li}, \citenamefont {Sit},\ and\ \citenamefont
  {Selloni}}]{chen2013chemical}%
  \BibitemOpen
  \bibfield  {author} {\bibinfo {author} {\bibfnamefont {J.}~\bibnamefont
  {Chen}}, \bibinfo {author} {\bibfnamefont {Y.-F.}\ \bibnamefont {Li}},
  \bibinfo {author} {\bibfnamefont {P.}~\bibnamefont {Sit}}, \ and\ \bibinfo
  {author} {\bibfnamefont {A.}~\bibnamefont {Selloni}},\ }\href@noop {}
  {\bibfield  {journal} {\bibinfo  {journal} {Journal of the American Chemical
  Society}\ }\textbf {\bibinfo {volume} {135}},\ \bibinfo {pages} {18774}
  (\bibinfo {year} {2013})}\BibitemShut {NoStop}%
\bibitem [{\citenamefont {Migani}\ \emph {et~al.}(2013)\citenamefont {Migani},
  \citenamefont {Mowbray}, \citenamefont {Iacomino}, \citenamefont {Zhao},
  \citenamefont {Petek},\ and\ \citenamefont {Rubio}}]{migani2013level}%
  \BibitemOpen
  \bibfield  {author} {\bibinfo {author} {\bibfnamefont {A.}~\bibnamefont
  {Migani}}, \bibinfo {author} {\bibfnamefont {D.~J.}\ \bibnamefont {Mowbray}},
  \bibinfo {author} {\bibfnamefont {A.}~\bibnamefont {Iacomino}}, \bibinfo
  {author} {\bibfnamefont {J.}~\bibnamefont {Zhao}}, \bibinfo {author}
  {\bibfnamefont {H.}~\bibnamefont {Petek}}, \ and\ \bibinfo {author}
  {\bibfnamefont {A.}~\bibnamefont {Rubio}},\ }\href@noop {} {\bibfield
  {journal} {\bibinfo  {journal} {Journal of the American Chemical Society}\
  }\textbf {\bibinfo {volume} {135}},\ \bibinfo {pages} {11429} (\bibinfo
  {year} {2013})}\BibitemShut {NoStop}%
\bibitem [{\citenamefont {Hao}\ \emph {et~al.}(2017)\citenamefont {Hao},
  \citenamefont {Wang}, \citenamefont {Dai}, \citenamefont {Zhou},\ and\
  \citenamefont {Yang}}]{hao2018photoelectron}%
  \BibitemOpen
  \bibfield  {author} {\bibinfo {author} {\bibfnamefont {Q.-q.}\ \bibnamefont
  {Hao}}, \bibinfo {author} {\bibfnamefont {Z.-q.}\ \bibnamefont {Wang}},
  \bibinfo {author} {\bibfnamefont {D.-x.}\ \bibnamefont {Dai}}, \bibinfo
  {author} {\bibfnamefont {C.-y.}\ \bibnamefont {Zhou}}, \ and\ \bibinfo
  {author} {\bibfnamefont {X.-m.}\ \bibnamefont {Yang}},\ }\href@noop {}
  {\bibfield  {journal} {\bibinfo  {journal} {Chinese Journal of Chemical
  Physics}\ }\textbf {\bibinfo {volume} {30}},\ \bibinfo {pages} {626}
  (\bibinfo {year} {2017})}\BibitemShut {NoStop}%
\bibitem [{\citenamefont {Migani}\ \emph {et~al.}(2014)\citenamefont {Migani},
  \citenamefont {Mowbray}, \citenamefont {Zhao}, \citenamefont {Petek},\ and\
  \citenamefont {Rubio}}]{migani2014quasiparticle}%
  \BibitemOpen
  \bibfield  {author} {\bibinfo {author} {\bibfnamefont {A.}~\bibnamefont
  {Migani}}, \bibinfo {author} {\bibfnamefont {D.~J.}\ \bibnamefont {Mowbray}},
  \bibinfo {author} {\bibfnamefont {J.}~\bibnamefont {Zhao}}, \bibinfo {author}
  {\bibfnamefont {H.}~\bibnamefont {Petek}}, \ and\ \bibinfo {author}
  {\bibfnamefont {A.}~\bibnamefont {Rubio}},\ }\href@noop {} {\bibfield
  {journal} {\bibinfo  {journal} {Journal of chemical theory and computation}\
  }\textbf {\bibinfo {volume} {10}},\ \bibinfo {pages} {2103} (\bibinfo {year}
  {2014})}\BibitemShut {NoStop}%
\bibitem [{\citenamefont {Di~Valentin}\ and\ \citenamefont
  {Fittipaldi}(2013)}]{di2013hole}%
  \BibitemOpen
  \bibfield  {author} {\bibinfo {author} {\bibfnamefont {C.}~\bibnamefont
  {Di~Valentin}}\ and\ \bibinfo {author} {\bibfnamefont {D.}~\bibnamefont
  {Fittipaldi}},\ }\href@noop {} {\bibfield  {journal} {\bibinfo  {journal}
  {The journal of physical chemistry letters}\ }\textbf {\bibinfo {volume}
  {4}},\ \bibinfo {pages} {1901} (\bibinfo {year} {2013})}\BibitemShut
  {NoStop}%
\bibitem [{\citenamefont {Ji}\ \emph {et~al.}(2014)\citenamefont {Ji},
  \citenamefont {Wang},\ and\ \citenamefont {Luo}}]{ji2014comparative}%
  \BibitemOpen
  \bibfield  {author} {\bibinfo {author} {\bibfnamefont {Y.}~\bibnamefont
  {Ji}}, \bibinfo {author} {\bibfnamefont {B.}~\bibnamefont {Wang}}, \ and\
  \bibinfo {author} {\bibfnamefont {Y.}~\bibnamefont {Luo}},\ }\href@noop {}
  {\bibfield  {journal} {\bibinfo  {journal} {The Journal of Physical Chemistry
  C}\ }\textbf {\bibinfo {volume} {118}},\ \bibinfo {pages} {21457} (\bibinfo
  {year} {2014})}\BibitemShut {NoStop}%
\bibitem [{\citenamefont {Di~Valentin}(2016)}]{di2016mechanism}%
  \BibitemOpen
  \bibfield  {author} {\bibinfo {author} {\bibfnamefont {C.}~\bibnamefont
  {Di~Valentin}},\ }\href@noop {} {\bibfield  {journal} {\bibinfo  {journal}
  {Journal of Physics: Condensed Matter}\ }\textbf {\bibinfo {volume} {28}},\
  \bibinfo {pages} {074002} (\bibinfo {year} {2016})}\BibitemShut {NoStop}%
\bibitem [{\citenamefont {Cheng}\ and\ \citenamefont
  {Sprik}(2012)}]{cheng2012alignment}%
  \BibitemOpen
  \bibfield  {author} {\bibinfo {author} {\bibfnamefont {J.}~\bibnamefont
  {Cheng}}\ and\ \bibinfo {author} {\bibfnamefont {M.}~\bibnamefont {Sprik}},\
  }\href@noop {} {\bibfield  {journal} {\bibinfo  {journal} {Physical Chemistry
  Chemical Physics}\ }\textbf {\bibinfo {volume} {14}},\ \bibinfo {pages}
  {11245} (\bibinfo {year} {2012})}\BibitemShut {NoStop}%
\bibitem [{\citenamefont {Cheng}\ \emph
  {et~al.}(2014{\natexlab{a}})\citenamefont {Cheng}, \citenamefont
  {VandeVondele},\ and\ \citenamefont {Sprik}}]{cheng2014identifying}%
  \BibitemOpen
  \bibfield  {author} {\bibinfo {author} {\bibfnamefont {J.}~\bibnamefont
  {Cheng}}, \bibinfo {author} {\bibfnamefont {J.}~\bibnamefont {VandeVondele}},
  \ and\ \bibinfo {author} {\bibfnamefont {M.}~\bibnamefont {Sprik}},\
  }\href@noop {} {\bibfield  {journal} {\bibinfo  {journal} {The Journal of
  Physical Chemistry C}\ }\textbf {\bibinfo {volume} {118}},\ \bibinfo {pages}
  {5437} (\bibinfo {year} {2014}{\natexlab{a}})}\BibitemShut {NoStop}%
\bibitem [{\citenamefont {Cheng}\ \emph {et~al.}(2015)\citenamefont {Cheng},
  \citenamefont {Liu}, \citenamefont {VandeVondele},\ and\ \citenamefont
  {Sprik}}]{cheng2015reductive}%
  \BibitemOpen
  \bibfield  {author} {\bibinfo {author} {\bibfnamefont {J.}~\bibnamefont
  {Cheng}}, \bibinfo {author} {\bibfnamefont {X.}~\bibnamefont {Liu}}, \bibinfo
  {author} {\bibfnamefont {J.}~\bibnamefont {VandeVondele}}, \ and\ \bibinfo
  {author} {\bibfnamefont {M.}~\bibnamefont {Sprik}},\ }\href@noop {}
  {\bibfield  {journal} {\bibinfo  {journal} {Electrochimica Acta}\ }\textbf
  {\bibinfo {volume} {179}},\ \bibinfo {pages} {658} (\bibinfo {year}
  {2015})}\BibitemShut {NoStop}%
\bibitem [{\citenamefont {Cheng}\ \emph
  {et~al.}(2014{\natexlab{b}})\citenamefont {Cheng}, \citenamefont {Liu},
  \citenamefont {Kattirtzi}, \citenamefont {VandeVondele},\ and\ \citenamefont
  {Sprik}}]{cheng2014aligning}%
  \BibitemOpen
  \bibfield  {author} {\bibinfo {author} {\bibfnamefont {J.}~\bibnamefont
  {Cheng}}, \bibinfo {author} {\bibfnamefont {X.}~\bibnamefont {Liu}}, \bibinfo
  {author} {\bibfnamefont {J.~A.}\ \bibnamefont {Kattirtzi}}, \bibinfo {author}
  {\bibfnamefont {J.}~\bibnamefont {VandeVondele}}, \ and\ \bibinfo {author}
  {\bibfnamefont {M.}~\bibnamefont {Sprik}},\ }\href@noop {} {\bibfield
  {journal} {\bibinfo  {journal} {Angewandte Chemie International Edition}\
  }\textbf {\bibinfo {volume} {53}},\ \bibinfo {pages} {12046} (\bibinfo {year}
  {2014}{\natexlab{b}})}\BibitemShut {NoStop}%
\bibitem [{\citenamefont {Hammes-Schiffer}\ and\ \citenamefont
  {Stuchebrukhov}(2010)}]{hammes2010theory}%
  \BibitemOpen
  \bibfield  {author} {\bibinfo {author} {\bibfnamefont {S.}~\bibnamefont
  {Hammes-Schiffer}}\ and\ \bibinfo {author} {\bibfnamefont {A.~A.}\
  \bibnamefont {Stuchebrukhov}},\ }\href@noop {} {\bibfield  {journal}
  {\bibinfo  {journal} {Chemical reviews}\ }\textbf {\bibinfo {volume} {110}},\
  \bibinfo {pages} {6939} (\bibinfo {year} {2010})}\BibitemShut {NoStop}%
\bibitem [{\citenamefont {Ghosh}\ \emph {et~al.}(2017)\citenamefont {Ghosh},
  \citenamefont {Castillo-Lora}, \citenamefont {Soudackov}, \citenamefont
  {Mayer},\ and\ \citenamefont {Hammes-Schiffer}}]{ghosh2017theoretical}%
  \BibitemOpen
  \bibfield  {author} {\bibinfo {author} {\bibfnamefont {S.}~\bibnamefont
  {Ghosh}}, \bibinfo {author} {\bibfnamefont {J.}~\bibnamefont
  {Castillo-Lora}}, \bibinfo {author} {\bibfnamefont {A.~V.}\ \bibnamefont
  {Soudackov}}, \bibinfo {author} {\bibfnamefont {J.~M.}\ \bibnamefont
  {Mayer}}, \ and\ \bibinfo {author} {\bibfnamefont {S.}~\bibnamefont
  {Hammes-Schiffer}},\ }\href@noop {} {\bibfield  {journal} {\bibinfo
  {journal} {Nano letters}\ }\textbf {\bibinfo {volume} {17}},\ \bibinfo
  {pages} {5762} (\bibinfo {year} {2017})}\BibitemShut {NoStop}%
\bibitem [{\citenamefont {Hammes-Schiffer}\ and\ \citenamefont
  {Soudackov}(2008)}]{hammes2008proton}%
  \BibitemOpen
  \bibfield  {author} {\bibinfo {author} {\bibfnamefont {S.}~\bibnamefont
  {Hammes-Schiffer}}\ and\ \bibinfo {author} {\bibfnamefont {A.~V.}\
  \bibnamefont {Soudackov}},\ }\href@noop {} {\bibfield  {journal} {\bibinfo
  {journal} {The Journal of Physical Chemistry B}\ }\textbf {\bibinfo {volume}
  {112}},\ \bibinfo {pages} {14108} (\bibinfo {year} {2008})}\BibitemShut
  {NoStop}%
\bibitem [{\citenamefont {Georgievskii}\ and\ \citenamefont
  {Stuchebrukhov}(2000)}]{georgievskii2000concerted}%
  \BibitemOpen
  \bibfield  {author} {\bibinfo {author} {\bibfnamefont {Y.}~\bibnamefont
  {Georgievskii}}\ and\ \bibinfo {author} {\bibfnamefont {A.~A.}\ \bibnamefont
  {Stuchebrukhov}},\ }\href@noop {} {\bibfield  {journal} {\bibinfo  {journal}
  {The Journal of Chemical Physics}\ }\textbf {\bibinfo {volume} {113}},\
  \bibinfo {pages} {10438} (\bibinfo {year} {2000})}\BibitemShut {NoStop}%
\bibitem [{\citenamefont {Nitzan}(2006)}]{nitzan2006chemical}%
  \BibitemOpen
  \bibfield  {author} {\bibinfo {author} {\bibfnamefont {A.}~\bibnamefont
  {Nitzan}},\ }\href@noop {} {\emph {\bibinfo {title} {Chemical dynamics in
  condensed phases: relaxation, transfer and reactions in condensed molecular
  systems}}}\ (\bibinfo  {publisher} {Oxford university press},\ \bibinfo
  {year} {2006})\BibitemShut {NoStop}%
\bibitem [{\citenamefont {Kondov}\ \emph {et~al.}(2007)\citenamefont {Kondov},
  \citenamefont {{\v{C}}{\'\i}{\v{z}}ek}, \citenamefont {Benesch},
  \citenamefont {Wang},\ and\ \citenamefont {Thoss}}]{kondov2007quantum}%
  \BibitemOpen
  \bibfield  {author} {\bibinfo {author} {\bibfnamefont {I.}~\bibnamefont
  {Kondov}}, \bibinfo {author} {\bibfnamefont {M.}~\bibnamefont
  {{\v{C}}{\'\i}{\v{z}}ek}}, \bibinfo {author} {\bibfnamefont {C.}~\bibnamefont
  {Benesch}}, \bibinfo {author} {\bibfnamefont {H.}~\bibnamefont {Wang}}, \
  and\ \bibinfo {author} {\bibfnamefont {M.}~\bibnamefont {Thoss}},\
  }\href@noop {} {\bibfield  {journal} {\bibinfo  {journal} {The Journal of
  Physical Chemistry C}\ }\textbf {\bibinfo {volume} {111}},\ \bibinfo {pages}
  {11970} (\bibinfo {year} {2007})}\BibitemShut {NoStop}%
\bibitem [{\citenamefont {Futera}\ and\ \citenamefont
  {Blumberger}(2017)}]{futera2017electronic}%
  \BibitemOpen
  \bibfield  {author} {\bibinfo {author} {\bibfnamefont {Z.}~\bibnamefont
  {Futera}}\ and\ \bibinfo {author} {\bibfnamefont {J.}~\bibnamefont
  {Blumberger}},\ }\href@noop {} {\bibfield  {journal} {\bibinfo  {journal}
  {The Journal of Physical Chemistry C}\ }\textbf {\bibinfo {volume} {121}},\
  \bibinfo {pages} {19677} (\bibinfo {year} {2017})}\BibitemShut {NoStop}%
\bibitem [{\citenamefont {Fernandez}\ \emph {et~al.}(2012)\citenamefont
  {Fernandez}, \citenamefont {Horvath},\ and\ \citenamefont
  {Hammes-Schiffer}}]{fernandez2012theoretical}%
  \BibitemOpen
  \bibfield  {author} {\bibinfo {author} {\bibfnamefont {L.~E.}\ \bibnamefont
  {Fernandez}}, \bibinfo {author} {\bibfnamefont {S.}~\bibnamefont {Horvath}},
  \ and\ \bibinfo {author} {\bibfnamefont {S.}~\bibnamefont
  {Hammes-Schiffer}},\ }\href@noop {} {\bibfield  {journal} {\bibinfo
  {journal} {The Journal of Physical Chemistry C}\ }\textbf {\bibinfo {volume}
  {116}},\ \bibinfo {pages} {3171} (\bibinfo {year} {2012})}\BibitemShut
  {NoStop}%
\bibitem [{\citenamefont {Burdett}\ \emph {et~al.}(1987)\citenamefont
  {Burdett}, \citenamefont {Hughbanks}, \citenamefont {Miller}, \citenamefont
  {Richardson~Jr},\ and\ \citenamefont {Smith}}]{burdett1987structural}%
  \BibitemOpen
  \bibfield  {author} {\bibinfo {author} {\bibfnamefont {J.~K.}\ \bibnamefont
  {Burdett}}, \bibinfo {author} {\bibfnamefont {T.}~\bibnamefont {Hughbanks}},
  \bibinfo {author} {\bibfnamefont {G.~J.}\ \bibnamefont {Miller}}, \bibinfo
  {author} {\bibfnamefont {J.~W.}\ \bibnamefont {Richardson~Jr}}, \ and\
  \bibinfo {author} {\bibfnamefont {J.~V.}\ \bibnamefont {Smith}},\ }\href@noop
  {} {\bibfield  {journal} {\bibinfo  {journal} {Journal of the American
  Chemical Society}\ }\textbf {\bibinfo {volume} {109}},\ \bibinfo {pages}
  {3639} (\bibinfo {year} {1987})}\BibitemShut {NoStop}%
\bibitem [{\citenamefont {Cheng}\ and\ \citenamefont
  {Sprik}(2010{\natexlab{a}})}]{cheng2010acidity}%
  \BibitemOpen
  \bibfield  {author} {\bibinfo {author} {\bibfnamefont {J.}~\bibnamefont
  {Cheng}}\ and\ \bibinfo {author} {\bibfnamefont {M.}~\bibnamefont {Sprik}},\
  }\href@noop {} {\bibfield  {journal} {\bibinfo  {journal} {Journal of
  chemical theory and computation}\ }\textbf {\bibinfo {volume} {6}},\ \bibinfo
  {pages} {880} (\bibinfo {year} {2010}{\natexlab{a}})}\BibitemShut {NoStop}%
\bibitem [{\citenamefont {Cheng}\ and\ \citenamefont
  {Sprik}(2010{\natexlab{b}})}]{cheng2010aligning}%
  \BibitemOpen
  \bibfield  {author} {\bibinfo {author} {\bibfnamefont {J.}~\bibnamefont
  {Cheng}}\ and\ \bibinfo {author} {\bibfnamefont {M.}~\bibnamefont {Sprik}},\
  }\href@noop {} {\bibfield  {journal} {\bibinfo  {journal} {Physical Review
  B}\ }\textbf {\bibinfo {volume} {82}},\ \bibinfo {pages} {081406} (\bibinfo
  {year} {2010}{\natexlab{b}})}\BibitemShut {NoStop}%
\bibitem [{\citenamefont {Cheng}\ \emph {et~al.}(2012)\citenamefont {Cheng},
  \citenamefont {Sulpizi}, \citenamefont {VandeVondele},\ and\ \citenamefont
  {Sprik}}]{cheng2012hole}%
  \BibitemOpen
  \bibfield  {author} {\bibinfo {author} {\bibfnamefont {J.}~\bibnamefont
  {Cheng}}, \bibinfo {author} {\bibfnamefont {M.}~\bibnamefont {Sulpizi}},
  \bibinfo {author} {\bibfnamefont {J.}~\bibnamefont {VandeVondele}}, \ and\
  \bibinfo {author} {\bibfnamefont {M.}~\bibnamefont {Sprik}},\ }\href@noop {}
  {\bibfield  {journal} {\bibinfo  {journal} {ChemCatChem}\ }\textbf {\bibinfo
  {volume} {4}},\ \bibinfo {pages} {636} (\bibinfo {year} {2012})}\BibitemShut
  {NoStop}%
\bibitem [{\citenamefont {VandeVondele}\ \emph {et~al.}(2005)\citenamefont
  {VandeVondele}, \citenamefont {Krack}, \citenamefont {Mohamed}, \citenamefont
  {Parrinello}, \citenamefont {Chassaing},\ and\ \citenamefont
  {Hutter}}]{vandevondele2005quickstep}%
  \BibitemOpen
  \bibfield  {author} {\bibinfo {author} {\bibfnamefont {J.}~\bibnamefont
  {VandeVondele}}, \bibinfo {author} {\bibfnamefont {M.}~\bibnamefont {Krack}},
  \bibinfo {author} {\bibfnamefont {F.}~\bibnamefont {Mohamed}}, \bibinfo
  {author} {\bibfnamefont {M.}~\bibnamefont {Parrinello}}, \bibinfo {author}
  {\bibfnamefont {T.}~\bibnamefont {Chassaing}}, \ and\ \bibinfo {author}
  {\bibfnamefont {J.}~\bibnamefont {Hutter}},\ }\href@noop {} {\bibfield
  {journal} {\bibinfo  {journal} {Computer Physics Communications}\ }\textbf
  {\bibinfo {volume} {167}},\ \bibinfo {pages} {103} (\bibinfo {year}
  {2005})}\BibitemShut {NoStop}%
\bibitem [{\citenamefont {Hutter}\ \emph {et~al.}(2014)\citenamefont {Hutter},
  \citenamefont {Iannuzzi}, \citenamefont {Schiffmann},\ and\ \citenamefont
  {VandeVondele}}]{hutter2014cp2k}%
  \BibitemOpen
  \bibfield  {author} {\bibinfo {author} {\bibfnamefont {J.}~\bibnamefont
  {Hutter}}, \bibinfo {author} {\bibfnamefont {M.}~\bibnamefont {Iannuzzi}},
  \bibinfo {author} {\bibfnamefont {F.}~\bibnamefont {Schiffmann}}, \ and\
  \bibinfo {author} {\bibfnamefont {J.}~\bibnamefont {VandeVondele}},\
  }\href@noop {} {\bibfield  {journal} {\bibinfo  {journal} {Wiley
  Interdisciplinary Reviews: Computational Molecular Science}\ }\textbf
  {\bibinfo {volume} {4}},\ \bibinfo {pages} {15} (\bibinfo {year}
  {2014})}\BibitemShut {NoStop}%
\bibitem [{\citenamefont {Hartwigsen}\ \emph {et~al.}(1998)\citenamefont
  {Hartwigsen}, \citenamefont {G{\oe}decker},\ and\ \citenamefont
  {Hutter}}]{hartwigsen1998relativistic}%
  \BibitemOpen
  \bibfield  {author} {\bibinfo {author} {\bibfnamefont {C.}~\bibnamefont
  {Hartwigsen}}, \bibinfo {author} {\bibfnamefont {S.}~\bibnamefont
  {G{\oe}decker}}, \ and\ \bibinfo {author} {\bibfnamefont {J.}~\bibnamefont
  {Hutter}},\ }\href@noop {} {\bibfield  {journal} {\bibinfo  {journal}
  {Physical Review B}\ }\textbf {\bibinfo {volume} {58}},\ \bibinfo {pages}
  {3641} (\bibinfo {year} {1998})}\BibitemShut {NoStop}%
\bibitem [{\citenamefont {VandeVondele}\ and\ \citenamefont
  {Hutter}(2007)}]{vandevondele2007gaussian}%
  \BibitemOpen
  \bibfield  {author} {\bibinfo {author} {\bibfnamefont {J.}~\bibnamefont
  {VandeVondele}}\ and\ \bibinfo {author} {\bibfnamefont {J.}~\bibnamefont
  {Hutter}},\ }\href@noop {} {\bibfield  {journal} {\bibinfo  {journal} {The
  Journal of chemical physics}\ }\textbf {\bibinfo {volume} {127}},\ \bibinfo
  {pages} {114105} (\bibinfo {year} {2007})}\BibitemShut {NoStop}%
\bibitem [{\citenamefont {Kulik}(2015)}]{kulik2015perspective}%
  \BibitemOpen
  \bibfield  {author} {\bibinfo {author} {\bibfnamefont {H.~J.}\ \bibnamefont
  {Kulik}},\ }\href@noop {} {\bibfield  {journal} {\bibinfo  {journal} {The
  Journal of chemical physics}\ }\textbf {\bibinfo {volume} {142}},\ \bibinfo
  {pages} {240901} (\bibinfo {year} {2015})}\BibitemShut {NoStop}%
\bibitem [{\citenamefont {Wang}\ \emph {et~al.}(2015)\citenamefont {Wang},
  \citenamefont {Wang},\ and\ \citenamefont {Hu}}]{wang2015identifying}%
  \BibitemOpen
  \bibfield  {author} {\bibinfo {author} {\bibfnamefont {D.}~\bibnamefont
  {Wang}}, \bibinfo {author} {\bibfnamefont {H.}~\bibnamefont {Wang}}, \ and\
  \bibinfo {author} {\bibfnamefont {P.}~\bibnamefont {Hu}},\ }\href@noop {}
  {\bibfield  {journal} {\bibinfo  {journal} {Physical Chemistry Chemical
  Physics}\ }\textbf {\bibinfo {volume} {17}},\ \bibinfo {pages} {1549}
  (\bibinfo {year} {2015})}\BibitemShut {NoStop}%
\bibitem [{\citenamefont {Heyd}\ \emph {et~al.}(2003)\citenamefont {Heyd},
  \citenamefont {Scuseria},\ and\ \citenamefont {Ernzerhof}}]{heyd2003hybrid}%
  \BibitemOpen
  \bibfield  {author} {\bibinfo {author} {\bibfnamefont {J.}~\bibnamefont
  {Heyd}}, \bibinfo {author} {\bibfnamefont {G.~E.}\ \bibnamefont {Scuseria}},
  \ and\ \bibinfo {author} {\bibfnamefont {M.}~\bibnamefont {Ernzerhof}},\
  }\href@noop {} {\bibfield  {journal} {\bibinfo  {journal} {The Journal of
  chemical physics}\ }\textbf {\bibinfo {volume} {118}},\ \bibinfo {pages}
  {8207} (\bibinfo {year} {2003})}\BibitemShut {NoStop}%
\bibitem [{\citenamefont {Krukau}\ \emph {et~al.}(2006)\citenamefont {Krukau},
  \citenamefont {Vydrov}, \citenamefont {Izmaylov},\ and\ \citenamefont
  {Scuseria}}]{krukau2006influence}%
  \BibitemOpen
  \bibfield  {author} {\bibinfo {author} {\bibfnamefont {A.~V.}\ \bibnamefont
  {Krukau}}, \bibinfo {author} {\bibfnamefont {O.~A.}\ \bibnamefont {Vydrov}},
  \bibinfo {author} {\bibfnamefont {A.~F.}\ \bibnamefont {Izmaylov}}, \ and\
  \bibinfo {author} {\bibfnamefont {G.~E.}\ \bibnamefont {Scuseria}},\
  }\href@noop {} {\bibfield  {journal} {\bibinfo  {journal} {The Journal of
  chemical physics}\ }\textbf {\bibinfo {volume} {125}},\ \bibinfo {pages}
  {224106} (\bibinfo {year} {2006})}\BibitemShut {NoStop}%
\bibitem [{\citenamefont {Guidon}\ \emph {et~al.}(2010)\citenamefont {Guidon},
  \citenamefont {Hutter},\ and\ \citenamefont
  {VandeVondele}}]{guidon2010auxiliary}%
  \BibitemOpen
  \bibfield  {author} {\bibinfo {author} {\bibfnamefont {M.}~\bibnamefont
  {Guidon}}, \bibinfo {author} {\bibfnamefont {J.}~\bibnamefont {Hutter}}, \
  and\ \bibinfo {author} {\bibfnamefont {J.}~\bibnamefont {VandeVondele}},\
  }\href@noop {} {\bibfield  {journal} {\bibinfo  {journal} {Journal of
  chemical theory and computation}\ }\textbf {\bibinfo {volume} {6}},\ \bibinfo
  {pages} {2348} (\bibinfo {year} {2010})}\BibitemShut {NoStop}%
\bibitem [{\citenamefont {Guidon}\ \emph {et~al.}(2009)\citenamefont {Guidon},
  \citenamefont {Hutter},\ and\ \citenamefont
  {VandeVondele}}]{guidon2009robust}%
  \BibitemOpen
  \bibfield  {author} {\bibinfo {author} {\bibfnamefont {M.}~\bibnamefont
  {Guidon}}, \bibinfo {author} {\bibfnamefont {J.}~\bibnamefont {Hutter}}, \
  and\ \bibinfo {author} {\bibfnamefont {J.}~\bibnamefont {VandeVondele}},\
  }\href@noop {} {\bibfield  {journal} {\bibinfo  {journal} {Journal of
  chemical theory and computation}\ }\textbf {\bibinfo {volume} {5}},\ \bibinfo
  {pages} {3010} (\bibinfo {year} {2009})}\BibitemShut {NoStop}%
\bibitem [{\citenamefont {Guidon}\ \emph {et~al.}(2008)\citenamefont {Guidon},
  \citenamefont {Schiffmann}, \citenamefont {Hutter},\ and\ \citenamefont
  {VandeVondele}}]{guidon2008ab}%
  \BibitemOpen
  \bibfield  {author} {\bibinfo {author} {\bibfnamefont {M.}~\bibnamefont
  {Guidon}}, \bibinfo {author} {\bibfnamefont {F.}~\bibnamefont {Schiffmann}},
  \bibinfo {author} {\bibfnamefont {J.}~\bibnamefont {Hutter}}, \ and\ \bibinfo
  {author} {\bibfnamefont {J.}~\bibnamefont {VandeVondele}},\ }\href@noop {}
  {\bibfield  {journal} {\bibinfo  {journal} {The Journal of chemical physics}\
  }\textbf {\bibinfo {volume} {128}},\ \bibinfo {pages} {214104} (\bibinfo
  {year} {2008})}\BibitemShut {NoStop}%
\bibitem [{\citenamefont {Grimme}\ \emph {et~al.}(2010)\citenamefont {Grimme},
  \citenamefont {Antony}, \citenamefont {Ehrlich},\ and\ \citenamefont
  {Krieg}}]{grimme2010consistent}%
  \BibitemOpen
  \bibfield  {author} {\bibinfo {author} {\bibfnamefont {S.}~\bibnamefont
  {Grimme}}, \bibinfo {author} {\bibfnamefont {J.}~\bibnamefont {Antony}},
  \bibinfo {author} {\bibfnamefont {S.}~\bibnamefont {Ehrlich}}, \ and\
  \bibinfo {author} {\bibfnamefont {H.}~\bibnamefont {Krieg}},\ }\href@noop {}
  {\bibfield  {journal} {\bibinfo  {journal} {The Journal of chemical physics}\
  }\textbf {\bibinfo {volume} {132}},\ \bibinfo {pages} {154104} (\bibinfo
  {year} {2010})}\BibitemShut {NoStop}%
\bibitem [{\citenamefont {Henkelman}\ \emph {et~al.}(2000)\citenamefont
  {Henkelman}, \citenamefont {Uberuaga},\ and\ \citenamefont
  {J{\'o}nsson}}]{henkelman2000climbing}%
  \BibitemOpen
  \bibfield  {author} {\bibinfo {author} {\bibfnamefont {G.}~\bibnamefont
  {Henkelman}}, \bibinfo {author} {\bibfnamefont {B.~P.}\ \bibnamefont
  {Uberuaga}}, \ and\ \bibinfo {author} {\bibfnamefont {H.}~\bibnamefont
  {J{\'o}nsson}},\ }\href@noop {} {\bibfield  {journal} {\bibinfo  {journal}
  {The Journal of chemical physics}\ }\textbf {\bibinfo {volume} {113}},\
  \bibinfo {pages} {9901} (\bibinfo {year} {2000})}\BibitemShut {NoStop}%
\bibitem [{\citenamefont {Skone}\ \emph {et~al.}(2006)\citenamefont {Skone},
  \citenamefont {Soudackov},\ and\ \citenamefont
  {Hammes-Schiffer}}]{skone2006calculation}%
  \BibitemOpen
  \bibfield  {author} {\bibinfo {author} {\bibfnamefont {J.~H.}\ \bibnamefont
  {Skone}}, \bibinfo {author} {\bibfnamefont {A.~V.}\ \bibnamefont
  {Soudackov}}, \ and\ \bibinfo {author} {\bibfnamefont {S.}~\bibnamefont
  {Hammes-Schiffer}},\ }\href@noop {} {\bibfield  {journal} {\bibinfo
  {journal} {Journal of the American Chemical Society}\ }\textbf {\bibinfo
  {volume} {128}},\ \bibinfo {pages} {16655} (\bibinfo {year}
  {2006})}\BibitemShut {NoStop}%
\bibitem [{\citenamefont {Dahl}\ and\ \citenamefont
  {Springborg}(1988)}]{dahl1988morse}%
  \BibitemOpen
  \bibfield  {author} {\bibinfo {author} {\bibfnamefont {J.~P.}\ \bibnamefont
  {Dahl}}\ and\ \bibinfo {author} {\bibfnamefont {M.}~\bibnamefont
  {Springborg}},\ }\href@noop {} {\bibfield  {journal} {\bibinfo  {journal}
  {The Journal of chemical physics}\ }\textbf {\bibinfo {volume} {88}},\
  \bibinfo {pages} {4535} (\bibinfo {year} {1988})}\BibitemShut {NoStop}%
\bibitem [{\citenamefont {Blumberger}(2015)}]{blumberger2015recent}%
  \BibitemOpen
  \bibfield  {author} {\bibinfo {author} {\bibfnamefont {J.}~\bibnamefont
  {Blumberger}},\ }\href@noop {} {\bibfield  {journal} {\bibinfo  {journal}
  {Chemical reviews}\ }\textbf {\bibinfo {volume} {{\bf 115}}},\ \bibinfo
  {pages} {11191} (\bibinfo {year} {2015})}\BibitemShut {NoStop}%
\bibitem [{\citenamefont {L{\"o}wdin}(1950)}]{lowdin1950non}%
  \BibitemOpen
  \bibfield  {author} {\bibinfo {author} {\bibfnamefont {P.-O.}\ \bibnamefont
  {L{\"o}wdin}},\ }\href@noop {} {\bibfield  {journal} {\bibinfo  {journal}
  {The Journal of Chemical Physics}\ }\textbf {\bibinfo {volume} {18}},\
  \bibinfo {pages} {365} (\bibinfo {year} {1950})}\BibitemShut {NoStop}%
\bibitem [{\citenamefont {Mayer}(2002)}]{mayer2002lowdin}%
  \BibitemOpen
  \bibfield  {author} {\bibinfo {author} {\bibfnamefont {I.}~\bibnamefont
  {Mayer}},\ }\href@noop {} {\bibfield  {journal} {\bibinfo  {journal}
  {International Journal of Quantum Chemistry}\ }\textbf {\bibinfo {volume}
  {90}},\ \bibinfo {pages} {63} (\bibinfo {year} {2002})}\BibitemShut {NoStop}%
\bibitem [{\citenamefont {Piela}(2013)}]{piela2013ideas}%
  \BibitemOpen
  \bibfield  {author} {\bibinfo {author} {\bibfnamefont {L.}~\bibnamefont
  {Piela}},\ }\href@noop {} {\emph {\bibinfo {title} {Ideas of quantum
  chemistry}}}\ (\bibinfo  {publisher} {Elsevier},\ \bibinfo {year}
  {2013})\BibitemShut {NoStop}%
\end{thebibliography}%

\end{document}